\documentclass[useAMS]{mn2e}
\usepackage[pdftex]{graphicx}
\pdfoutput=1

\usepackage{graphicx}

\title[C, N, O, and Na Abundances of Cepheid Variables]
{C, N, O, and Na Abundances of Cepheid Variables:\\
Implications on the Mixing Process in the Envelope\footnotemark[0]\thanks{
Based on observations carried out at Bohyunsan Astronomical Observatory 
of Korean Astronomy and Space Science Institute (KASI).}
\thanks{Large data of electronic tables are provided as supplementary 
materials.}
}

\author[Y. Takeda, D.-I. Kang, I. Han, B.-C. Lee, and K.-M. Kim]
{Y. Takeda$^{1}$\thanks{E-mail:
takeda.yoichi@nao.ac.jp}, 
D.-I. Kang,$^{2}$ I. Han,$^{3}$
B.-C. Lee,$^{3}$ and K.-M. Kim$^{3}$\\
$^{1}$National Astronomical Observatory of Japan, 
2-21-1 Osawa, Mitaka, Tokyo 181-8588, Japan\\
$^{2}$Changwon Science high school,
30, Pyungsanro 159-th, Uichang, Changwon, 641-500, Korea\\
$^{3}$Korea Astronomy and Space Science Institute,
61-1 Whaam-dong, Youseong-gu, Taejon 305-348, Korea
}
\begin{document}

\date{Accepted 2013 March 19. Received 2013 March 19; in original form 2013 February 19}


\maketitle

\label{firstpage}

\begin{abstract}
With an aim of investigating the nature of evolution-induced mixing 
in the envelope of evolved intermediate-mass stars, we carried out 
an extensive spectroscopic study for 12 Cepheid variables of various 
pulsation periods ($\sim $~2--16~days) to determine the photospheric 
abundances of C, N, O, and Na, which are the key elements for
investigating how the H-burning products are salvaged from the interior, 
based on 122 high-dispersion echelle spectra ($\sim 10$ per target) 
of wide wavelength coverage collected at Bohyunsan Astronomical Observatory. 
Having established the relevant atmospheric parameters corresponding 
to each phase spectroscopically from the equivalent widths of Fe~{\sc i} 
and Fe~{\sc ii} lines, we derived C, N, O, and Na abundances from 
C~{\sc i} 7111/7113/7115/7116/7119, O~{\sc i} 6155--8, 
N~{\sc i} 8680/8683/8686, and Na~{\sc i} 6154/6161 lines 
by using the spectrum-synthesis fitting technique, while taking 
into account the non-LTE effect.
The resulting abundances of these elements for 12 program stars 
turned out to show remarkably small star-to-star dispersions 
($\la$~0.1--0.2dex) without any significant dependence upon the 
pulsation period: near-solar Fe ([Fe/H]~$\sim 0.0$), moderately 
underabundant C ([C/H]~$\sim -0.3$), appreciably overabundant N 
([N/H]~$\sim$~+0.4--0.5), and mildly supersolar Na ([Na/H]~$\sim +0.2$). 
We conclude the following implications from these observational facts:
(1) These CNO abundance trends can be interpreted mainly as due to 
the canonical dredge-up of CN-cycled material, while any significant
non-canonical deep mixing of ON-cycled gas is ruled out (though only 
a slight mixing may still be possible). (2) The mild but definite 
overabundance of Na suggests that the NeNa-cycle product is also 
dredged up. 
(3) The extent of mixing-induced peculiarities in the envelope
of Cepheid variables is essentially independent on the absolute 
magnitude; i.e., also on the stellar mass. 
\end{abstract}

\begin{keywords}
stars: abundances -- stars: atmospheres --  stars: evolution
-- stars: variables: Cepheids --
stars: individual (SU~Cas, SZ~Tau, RT~Aur, $\zeta$~Gem, FF~Aql, 
$\eta$~Aql, S~Sge, X~Cyg, T~Vul, DT~Cyg, V1334~Cyg, $\delta$~Cep)
\end{keywords}

\section{Introduction}

As a star is evolved off the main sequence after exhaustion of
hydrogen fuels in the core, it increases its radius while the 
surface temperature drops down, and the deep convection zone is 
developed, by which part of the nuclear-processed products
in the core may be salvaged and mixed into the outer envelope 
and abundance peculiarities may be observed for some specific 
elements. By making use of this fact, we can study the physical 
process in the invisible interior of stars by comparing 
spectroscopically determined surface abundances with theoretical 
expectations from stellar evolution calculations.

While considerable progress has been made so far in this field
and important observational characteristics are known to be 
successfully explained by theoretical calculations, discrepancies 
between the empirically established surface abundances and the 
prediction from the standard theory are occasionally seen.
One of such problems concerns the oxygen abundances in F--G 
supergiants, which is namely the disagreement between the 
theoretical prediction (almost normal O) and an apparent 
deficit often suggested from observations.

In the envelope of such intermediate-mass supergiants, the 
H-burning (CNO cycle) products are mixed to alter the surface
abundances. According to the canonical stellar evolution
calculations (e.g., Lejeune \& Schaerer 2001), it is essentially 
the CN-cycled (C$\rightarrow$N reaction) material that is dredged up, 
while the product of ON-cycle (O$\rightarrow$N reaction; occurring 
in deeper region of higher $T$) is unlikely to cause any significant
abundance change because mixing is not expected to substantially 
penetrate into such a deep layer; 
thus the predicted surface abundances are characterized by 
a deficit in C as well as an enhancement in N, 
while O is practically unaffected (if only slightly decreased).

On the observational side, however, Luck \& Lambert (1981, 1985) 
reported based on the CNO abundance results of FGK supergiants
(including Cepheid variables) that oxygen is mildly deficient 
relative to the Sun by $\sim$~0.2--0.4~dex, while C is underabundant 
by $\sim$~0.2--1.0~dex and nitrogen is overabundant by 
$\sim$~0.2--0.8~dex (both with a large star-to-star dispersion).
In such a deficiency of O is real, a non-canonical deep mixing 
may exist that dredges the ON-cycle product up into the envelope.
Yet, Luck \& Lambert (1985) did not stick to this solution and 
discussed several possibilities (e.g., errors in abundance 
determinations, peculiarities in solar abundances, etc.).
Their tentative conclusion was that their absolute abundance 
determinations may not necessarily be trustworthy, pointing out 
that the trends of more reliable C/O ratios (from forbidden lines) 
as well as C/N ratios (from permitted lines) could be explained 
by canonical dredge-up of CN-cycled material (though its amount 
may be more severe than expected). Thus, the problem remained 
as an open question.

Meanwhile, Takeda and Takada-Hidai published a series of papers 
in 1990s (Takeda \& Takada-Hidai 1994, 1995, 1998, 2000), where 
the abundance peculiarities of supergiants in terms of Na, N, O, 
and C were investigated by taking into account the non-LTE effect,
though they did not pay much attention to determination of 
atmospheric parameters (appropriate values corresponding to the 
spectral type and the luminosity class were simply assumed).  
One of the important results they first learned was the 
confirmation of Na enrichment (Takeda \& Takada-Hidai 1994), 
which is regarded as an evidence for the dredge-up of NeNa-cycle 
product (cf. Sasselov 1986, Lambert 1992).
Since such a peculiarity in Na was not expected from canonical 
stellar evolution calculations, it implied an instructive fact 
that the standard theory was not necessarily sufficient.

Regarding CNO, Takeda \& Takada-Hidai (1998) confirmed the mild 
underabundance of O by $\sim 0.3$~dex, interestingly just like 
Luck \& Lambert (1985) reported, along with the enrichment of N 
(Takeda \& Takada-Hidai 1995) as well as the deficit of C 
(Takeda \& Takadai-Hidai 2000; though only for late-B through 
late-A supergiants). But Takeda \& Takada-Hidai (1998) did not 
associate this result of subsolar O with non-canonical mixing of 
ON-cycled material but attributed it to the general tendency of 
apparently subsolar CNO observed in young stars such as 
B-type stars (e.g., Nissen 1993), which was generally accepted 
(though puzzled) at that time.
However, an extensive non-LTE study on the oxygen abundances of 
B-type stars in comparison with the Sun recently carried out by
Takeda et al. (2010) lead to the conclusion that B stars have 
almost the solar composition in terms of O without any significant 
difference. If so, these previous studies may point to the conclusion 
that O is apparently underabundant in supergiants while primordial O 
must have been almost normal when they were born. Does this 
imply that oxygen abundance in the envelope of intermediate-mass
stars really got decreased during the course of evolution 
by a non-standard deep mixing of ON-cycled product? 

In the meantime, Kovtyukh \& Andrievsky (1999) suggested 
a possible key to the solution of this oxygen problem. 
According to them, the atmospheric parameters adopted by previous 
investigations may not have been appropriate, since spectroscopically 
determined parameters based on Fe~{\sc i} and Fe~{\sc ii} lines 
tend to suffer to appreciable errors because of the non-LTE 
overionization affect (particularly important for stronger 
Fe~{\sc i} lines). As a method to alleviate this problem, they 
proposed to principally invoke Fe~{\sc ii} lines (considered to 
be hardly affected by any non-LTE effect) of various strengths 
for determination of spectroscopic parameters 
while using only ``fairly weak'' Fe~{\sc i} lines 
to define the reference abundance derived from neutral Fe lines 
(which is to be made consistent with that from once-ionized Fe 
lines for the requirement of ionization equilibrium) as the 
weak-line limit under the assumption that very weak deep-forming 
Fe~{\sc i} lines are formed nearly in LTE. They tested this method
by applying it to analyzing the spectra of $\delta$ Cep at seven
different phases, and obtained quite consistent results between the 
such derived spectroscopic gravity and the physical gravity 
(determined from mass and radius), which eventually solved the 
long-standing gravity-discrepancy problem in Cepheids (e.g., 
Luck \& Lambert 1985). Then, as an important consequence, the 
resulting oxygen abundance with the new parameters (higher $\log g$) 
turned out to be nearly normal, in contrast to the case of old
parameters (lower $\log g$; derived by the conventional approach)
where an appreciable underabundance of oxygen was obtained.
So, if their claim is correct, the oxygen problem in F--G supergiants
may be nothing but an apparent effect due to an inadequate choice of 
atmospheric parameters; and the ``actual'' surface CNO abundances 
would be characterized by low C, high N, and nearly normal O, which 
may be reasonably explained by the theory of canonical mixing
(i.e., essentially the dredge-up of CN-cycle product). 

Motivated by the argument of Kovtyukh \& Andrievsky (1999), we decided 
to carry out an extensive spectroscopic analysis on the abundance 
peculiarities of C, N, O, and Na (the key elements for studying the 
mixing of H-burning products) for Cepheid variables of various 
pulsation periods based on high-dispersion spectra taken at different 
phases, in order to see whether their conclusion can be generally 
confirmed for evolved intermediate-mass stars of diversified properties, 
which is the purpose of this study.
The reason why we have chosen Cepheids (instead of non-variable
supergiants) is because (1) we can directly estimate the uncertainties 
in abundance determinations by comparing the results from spectra of 
different phases, and (2) the stellar parameters (such as absolute 
magnitude, mass, radius, etc.) are well known in advance by making 
use of the period--luminosity relation.

What we want to check in this investigation are the following points :\\
--- Are we able to derive reliable atmospheric parameters 
spectroscopically based on Fe~{\sc i} and Fe~{\sc ii} lines 
by the way proposed by Kovtyukh \& Andrievsky (1999)?
Can we observe an agreement between the spectroscopic 
gravity and the dynamical gravity?\\
--- Are the abundances derived from spectra corresponding to 
various phases reasonably consistent with each other?\\
--- What about the tendency of C, N, O, and Na abundances
in view of the specific trends resulting from mixing of 
H-burning products? In particular, how is the abundance of 
oxygen? Can we confirm that the sum of C+N+O is conserved?\\
--- Is there any meaningful dependence in the extent of abundance 
peculiarities upon the stellar mass (or luminosity directly related 
to the pulsation period), such as suggested previously (e.g., 
Takeda \& Takada-Hidai 1994) for the case of Na?

It may be worth noting here that extensive spectroscopic studies of Cepheid 
variables at various phases similar to ours were already conducted by 
Andrievsky, Luck \& Kovtyukh (2005), Luck \& Andrievsky (2004),
Kovtyukh et al. (2005a), and Luck et al. (2008), 
 for four different Cepheid groups 
of (i) $3 < P < 6$, (ii) $6 < P < 10$, 
(iii) $10 < P $ (where $P$ is the pulsation period in day), 
and (iv) s-Cepheids with small amplitudes, respectively, in order to study
the phase-dependent variations of stellar fundamental parameters.
They showed that almost consistent abundances were obtained
for various elements by using the spectroscopic parameters determined 
based on Kovtyukh \& Andrievsky's (2005) procedure.

In contrast, we treat and discuss our targets (including classical 
Cepheids in a wide range of $P$ as well as s-Cepheids) as a whole without 
classifying them into subgroups; but we particularly focus on determining 
the abundances of four elements (C, N, O, and Na) as precisely as possible 
by carefully applying the spectrum-synthesis technique along with 
non-LTE corrections, which is the distinction compared to their studies.

The remainder of this paper is organized as follows.
We describe in Section 2 the observational data of the program stars 
(12 Cepheid variables). The spectroscopic determination 
of atmospheric parameters corresponding to each spectrum, which uses 
Fe~{\sc i} and Fe~{\sc ii} lines, is explained in Section 3. 
In Section 4 are spelled out the procedures of our abundance analysis, 
which are made up of LTE spectrum-synthesis fitting, inverse evaluation 
of equivalent widths, and non-LTE abundance determinations from them. 
The results of atmospheric parameters as well as C, N, O, and Na 
abundances are discussed in Section 5 with respect to each of the
specific check points enumerated above. The conclusion is summarized
in Section 6.

\section{Observational Data}

Twelve representative Cepheid variables (including five s-Cepheids
showing particularly small light-variation amplitude of
$|\Delta V| \la 0.5$~mag) were selected as our program 
stars, which are apparently bright (mostly $V\sim$~4--6~mag) and have
a wide variety of pulsation periods ($\sim$2--16~days).
The basic data of these targets are summarized in Table 1.

The observations were carried out during the four nights on 2009 
October 2--5 by using BOES (Bohyunsan Observatory Echelle Spectrograph)
attached to the 1.8 m reflector at Bohyunsan Optical Astronomy Observatory. 
Using 2k$\times$4k CCD (pixel size of 15~$\mu$m~$\times$~15~$\mu$m), 
this echelle spectrograph enabled us to obtain spectra of wide 
wavelength coverage (from $\sim$~3800~$\rm\AA$ to $\sim$~9200~$\rm\AA$).
We used 200$\mu$m fiber corresponding to the resolving power 
of $R \simeq 45000$. The integration time for each exposure was 
from a few minutes up to $\sim$~15--20~minutes depending on the 
brightness of a target. If ADU counts attained by one exposure 
were not sufficient, the second (or even third) exposure was tried 
for the same star and the successive frames were co-added, though 
only one exposure was enough in normal cases. In any case, the 
integrated exposure time for each spectrum (whichever single or 
co-added) never exceeded 40 minutes, which guarantees that any
serious blurring effect caused by line-profile variations is 
unlikely. In such a way, each of the program stars 
were observed several times in a night with an interval of a few hours,
though the actual frequency differed from star to star.
Thus, as a result of 4-night observations, we could obtain a total of 
122 spectra, which consist of 7--17 spectra per each star corresponding 
to different observational times. 

The reduction of the echelle spectra (bias subtraction, flat 
fielding, spectrum extraction, wavelength calibration, and 
continuum normalization) was carried out mainly with the software 
developed by Kang et al. (2006) and partly with IRAF.\footnote{
IRAF is distributed by the National Optical Astronomy Observatories,
which is operated by the Association of Universities for Research
in Astronomy, Inc. under cooperative agreement with the National 
Science Foundation.} 
We could accomplish sufficiently high S/N ratio of several hundred 
at the relevant regions for most of the spectra (except for the 
last one of $\delta$~Cep, for which the quality is considerably poor).
Then, the apparent stellar radial velocity was determined by comparing 
each spectrum in the orange region ($\lambda \sim$~6100~$\rm\AA$) with the 
template solar spectrum, which was further converted to the heliocentric 
radial velocity by applying the relevant correction evaluated with the
help of IRAF task ``rvcorrect.''  
The fundamental information for each of the 122 spectra (observational 
time in Julian day, the corresponding pulsation phase $\phi$, and
the heliocentric radial velocity $V_{\rm rad}$) is presented in Table 2.

\section{Atmospheric Parameters}

The atmospheric parameters necessary for constructing model 
atmospheres as well as for abundance determinations [$T_{\rm eff}$ 
(effective temperature), $\log g$ (surface gravity), $\xi$ 
(microturbulent velocity dispersion), and [Fe/H] (metallicity, 
represented by the Fe abundance relative to the Sun)] were 
spectroscopically determined from the equivalent widths 
($W_{\lambda}$) of Fe~{\sc i} and Fe~{\sc ii} lines, based on the 
conventional requirements of (a) excitation equilibrium (Fe abundances 
show no systematic dependence on the excitation potential), (b)
ionization equilibrium (mean Fe abundance from Fe~{\sc i} lines 
and that from Fe~{\sc ii} lines agree with each other), and (c)
curve-of-growth matching (Fe abundances do not systematically 
depend on $W_{\lambda}$).

Practically, we used the program TGVIT (an updated version of 
the original program named TGV) developed for this purpose 
(Takeda et al. 2005; cf. section 3.1 therein), which can establish 
these four parameters simultaneously by numerically finding the 
best solution such that minimizing the dispersion function $D^{2}$
defined as the combination of $\langle \log \epsilon_{1} \rangle$ 
(mean Fe abundance derived from Fe~{\sc i} lines),
$\langle \log \epsilon_{2} \rangle$ 
(mean Fe abundance derived from Fe~{\sc ii} lines),
$\sigma_{1}$ (standard deviation of the Fe abundances from Fe~{\sc i} 
lines), and $\sigma_{2}$ (standard deviation of the Fe abundances 
from Fe~{\sc ii} lines) as
\begin{equation}
D^{2} \equiv (\sigma_{1}^{2} + \sigma_{2}^{2}) + 
(\langle \log \epsilon_{1} \rangle - 
\langle \log \epsilon_{2} \rangle)^2.
\end{equation}
The details regarding the principle and algorithm of this method
are described in Takeda, Ohkubo \& Sadakane (2002).

Regarding the observational $W_{\lambda}$ data of Fe lines, we could 
measure $W_{\lambda}$'s of $\sim$~100--250 Fe~{\sc i} lines and 
$\sim$~15--25 Fe~{\sc ii} lines on each of the 122 spectra (the number 
of measurable lines significantly depends on the spectral line 
width determined by macroscopic line broadening as well as on the 
S/N ratio), while consulting the line list (comprising 302 
Fe~{\sc i} lines and 28 Fe~{\sc ii} lines) given in electronic 
table E1 of Takeda et al. (2005), where Gaussian fitting was
adopted in most cases. In practice, avoiding the use of very 
strong lines showing appreciable damping wings, which are unsuitable 
for abundance determinations, we decided to use only those lines 
satisfying the condition of $w \le 200$~m$\rm\AA$,
where $w \equiv W_{\lambda} \cdot (5000/\lambda)$ is 
the reduced equivalent width to $\lambda = 5000$~$\rm\AA$.

We first carried out some test calculations by using the spectrum 
of $\delta$~Cep (213306-091002A), in order to confirm the argument 
of Kovtyukh \& Andrievsky (1999), who suggested that Fe~{\sc ii} 
lines of various strengths can be safely used, while the mean 
abundance from Fe~{\sc i} lines should be determined based on 
only weak lines (as the weak-line limit). Namely, the following 
two different cases were examined concerning the input data set
of $W_{\lambda}$.\\
--- Case (A):\\
The TGVIT program was applied without any specific constraint on 
the line strengths (i.e., all available Fe~{\sc i} and Fe~{\sc ii} 
lines satisfying $w \le 200$~m$\rm\AA$).
Then, the solutions of [$T_{\rm eff}$ (K), $\log g$ (cm~s$^{-1}$), 
$\xi$ (km~s$^{-1}$), and [Fe/H] (dex)] minimizing $D^{2}$ were 
obtained as [5694, 1.88, 3.56, and 0.00]. 
However, the abundances from Fe~{\sc i} and Fe~{\sc ii} lines 
corresponding to these parameters turned out appreciably 
$W_{\lambda}$-dependent in an opposite manner (i.e., 
Fe~{\sc i}~$|$~Fe~{\sc ii} abundances tend to progressively 
decrease~$|$~increase with an increase in $W_{\lambda}$) 
as shown in Fig. 1a, which is hardly acceptable.\\
--- Case (B):\\
Next, we restricted Fe~{\sc i} lines to only those fairly weak ones 
satisfying the criterion of $w \le 30$~m$\rm\AA$,
while Fe~{\sc ii} lines are usually used as before, in analogy 
with treatment suggested by Kovtyukh \& Andrievsky (1999).
The resulting parameter solutions for this case were
[5706, 2.19, 3.88, and +0.07]; the especially notable change 
compared to Case (A) is an increase of $\log g$ by $\sim$~0.3~dex.
The $\log \epsilon$ vs. $W_{\lambda}$ relation for this case
is displayed in Fig. 1c, where we can satisfactorily confirm
that the Fe abundances do not show any systematic dependence
upon $W_{\lambda}$. Naturally, as recognized in the $\log \epsilon$ 
vs. $\chi_{\rm low}$ plots, the abundance scatter around the 
mean is appreciably smaller for Case (B) (Fig. 1d) in comparison 
with Case (A) (Fig. 1b).

Given these results, which reconfirmed the consequence of 
Kovtyukh \& Andrievsky (1999), we decided to follow Case (B) and 
use only the weak Fe~{\sc i} lines with $w \le 30$~m$\rm\AA$
but all Fe~{\sc ii} lines with $w \le 200$~m$\rm\AA$,
as the input $W_{\lambda}$ data to which TGVIT was applied.
The number of actually used Fe~{\sc i} lines after this screening
was $\sim$~40--130 (i.e., almost as half as the originally measured ones).
The finally resulting parameters ($T_{\rm eff}$, $\log g$, 
$\xi$, and [Fe/H]) for each of the 122 spectra are presented in 
Table 2. The typical statistical uncertainties involved in these 
solutions estimated in the manner described in Section 3.2 of 
Takeda et al. (2002) are ($\sim 50$~K, $\sim 0.1$~dex, 
$\sim~0.3$~km~s$^{-1}$, and $\sim~0.05$~dex), respectively.
The corresponding $\log \epsilon$ vs. $W_{\lambda}$ and 
$\log \epsilon$ vs. $\chi_{\rm low}$ plots are depicted
in Fig. 2 and Fig.3, respectively, where we can see that 
the requirements (a), (b), and (c) are reasonably fulfilled.
The detailed line-by-line data ($W_{\lambda}$, $\log \epsilon$)
for each stellar spectrum are also given as the electronic
data tables contained in the supplementary materials (the results 
for HD~?????? are presented in the file named ``??????\_feabw.dat'').

\section{Abundance Determinations}

\subsection{Target Lines of C, N, O, and Na}

Regarding the choice of lines to be used for deriving the
abundances of C, N, O, and Na, several requirements
were taken into consideration:\\
--- From the viewpoint of mutual consistency, lines of the same type 
had better be used for all these elements; so we would rely on 
permitted lines of neutral species.\\
--- The strength of the line must not be too weak, 
so that firm detectability in the relevant parameter ranges 
($T_{\rm eff} \sim$~5000--7000~K, $\log g \sim$~1--3) 
can be guaranteed.\\
--- On the other hand, we would like to avoid considerably strong 
lines, such as those very sensitive to a non-LTE effect or 
microturbulence.\\
--- Since our analysis is based on the spectrum synthesis method,
it is preferable to select a wavelength region of moderate width
comprising a few lines belonging the same multiplet, 
so that they may be analyzed at a time.

We then decided to invoke the following lines which reasonably 
meet these conditions: 
C~{\sc i}~7111/7113/7115/7116/7119 lines for C, 
N~{\sc i}~8680/8683/8686 lines for N,
O~{\sc i}~6155--8 lines for O, and
Na~{\sc i} 6154/6161 lines for Na.

\subsection{Synthetic Spectrum Fitting}

Abundance determinations in the first step were carried out
by using our synthetic spectrum fitting code MPFIT, in which 
the spectrum synthesis part is originally based on Kurucz's 
(1993) WIDTH9 program. It accomplishes the best-match 
between the theoretical and observational spectra
based on the algorithm described in Takeda (1995),
by simultaneously varying the abundances of several key elements 
($\log\epsilon_{1}$, $\log\epsilon_{2}$, $\ldots$), macrobroadening 
parameter ($v_{\rm M}$), and the radial-velocity (wavelength) shift 
($\Delta \lambda$).
The macrobroadening parameter ($v_{\rm M}$) is the $e$-folding width
of the Gaussian macrobroadening function,
$M(v) \propto \exp[-(v/v_{\rm M})^{2}]$,
which represents the combined effects of instrumental broadening, 
macroturbulence, and rotational velocity (though it is 
essentially dominated by mactroturbulence in the present case). 

Practically, our spectrum fitting analysis was conducted for the 
following three wavelength regions, where the elements whose 
abundances were treated as variables are enumerated in each bracket: 
(i) 6143--6163~$\rm\AA$ [O, Na, Si, Ca, Fe] including 
O~{\sc i}~6155--8 and Na~{\sc i} 6154/6161 lines,
 (ii) 7110--7121~$\rm\AA$ [C, Fe, Ni] including 
C~{\sc i}~7111/7113/7115/7116/7119 lines
and (iii) 8677--8697~$\rm\AA$ [N, Si, S, Fe] 
including N~{\sc i}~8680/8683/8686 lines.

Regarding the atomic data of spectral lines (wavelengths, excitation
potentials, oscillator strengths, and damping constants), we basically 
invoked the compilation of Kurucz \& Bell (1995). However, pre-adjustments
of several $\log gf$ values were necessary (i.e., use of empirically 
determined solar $gf$ values) for several lines in 7110--7121~$\rm\AA$
region in order to accomplish a satisfactory match between the observed 
and theoretical spectrum. The finally adopted atomic data of 
important spectral lines are presented in table 2. 
The atmospheric model to be used for each spectrum was generated by 
interpolating Kurucz's (1993) grid of ATLAS9 model atmospheres in 
terms of $T_{\rm eff}$, $\log g$, and [Fe/H].

This fitting analysis turned out quite successful and we could obtain 
the abundances (especially for C, N, O, and Na) for all the 122 spectra.
In case where some part of the spectrum was found to be damaged due to 
cosmic rays or telluric lines (e.g., occasionally encountered at 
$\lambda \sim 7118$~$\rm\AA$ in the 7110--7121~$\rm\AA$ fitting), 
such a region was masked and excluded from the evaluation of 
$\chi^{2}(O-C)$ evaluation.
How the theoretical spectrum corresponding to the converged solutions 
fits well with the observed spectrum is displayed in 
figure 4 (6143--6163~$\rm\AA$ fitting), 
figure 5 (7110--7121~$\rm\AA$ fitting), and
figure 6 (8677--8697~$\rm\AA$  fitting). 
Note that we assumed LTE for all lines at this stage of synthetic 
spectrum-fitting.
The solutions of the macrobroadening  width ($v_{\rm M}$) derived from 
the 6143--6163~$\rm\AA$ fitting are given in Table 2, while further work 
is yet to be done to establish the final abundances as described in 
the next Section 4.3. 

\subsection{Equivalent Widths Analysis}

Despite that the synthetic spectrum fitting directly yielded the 
abundance solutions of C, N, O, and Na (the main purpose of this study), 
this approach is not necessarily suitable when one wants to examine 
how the results would be changed in different conditions
(e.g., LTE vs. non-LTE, application of non-LTE calculations done 
at different assumptions, abundance sensitivity to parameter changes, 
etc.), since it is tedious to repeat the fitting process to obtain
the new solution.
Therefore, with the help of the modified version of 
Kurucz's (1993) WIDTH9 program\footnote{
This WIDTH9 program had been considerably modified by Y. Takeda in various 
respects; e.g., inclusion of non-LTE effects, treatment of total 
equivalent width for multi-component lines; etc.}, 
we computed the equivalent widths for C~{\sc i} lines 
($W_{7111}$, $W_{7113}$, $W_{7115}$, $W_{7116}$, $W_{7119}$), 
N~{\sc i} lines ($W_{8680}$, $W_{8683}$, $W_{8686}$), 
O~{\sc i} lines ($W_{6155-8}$),\footnote{We use the total equivalent width
of the O~{\sc i} 6155--8 feature consisting of 9 component lines
(which appears as a merged triplet feature) in order to maintain
consistency with Takeda \& Takada-Hidai (1998) as well as 
Takeda et al. (2010). Since the actual O~{\sc i} 6155--8 feature is 
appreciably blended with lines of other elements (such as Si~{\sc i}, 
Ca~{\sc i}, and Fe~{\sc i} lines; cf. Fig. 4 and Table 3), this
$W_{6155-8}$) is not so much a realistic value measurable from actual spectra
as rather an idealized quantity.} and Na~{\sc i} lines 
($W_{6154}$, $W_{6161}$) ``inversely'' from the abundance solutions 
(resulting from spectrum synthesis) along with the adopted 
atmospheric model/parameters, which are much easier to handle.
Based on such evaluated $W_{\lambda}$ values, 
the LTE ($\log\epsilon^{\rm LTE}$) as well as non-LTE abundances 
($\log\epsilon^{\rm NLTE}$) were freshly computed, from which 
the non-LTE correction 
($\Delta [\equiv \log\epsilon^{\rm NLTE} - \log\epsilon^{\rm LTE}$])
was further derived.

Regarding the calculations for evaluating the non-LTE departure
coefficients, we followed the procedures described in Takeda
\& Takada-Hidai (2000) (for C), Takeda \& Takada-Hidai (1995) 
(for N),Takeda \& Takada-Hidai (1998) as well as Takeda et al. 
(2010) (for O), and Takeda \& Takada-Hidai (1994) (for Na), which 
should be consulted for the details.
In the actual non-LTE calculations for element X, two sets of
departure coefficients were prepared corresponding to two different 
input theoretical abundances ([X/H]$_{1}^{\rm t}$, [X/H]$_{2}^{\rm t}$; 
expressed as the values relative to the solar abundance)
which were so chosen as to encompass the real abundances of
the program stars; i.e., ($-0.5$, 0.0) for C, (0.0, +1.0) for N, 
(0.0, $-0.5$) for O, and (0.0, +0.5) for Na.
Then, from a given $W_{\lambda}$, we obtained two kinds of 
non-LTE abundances ($\log\epsilon_{1}$, $\log\epsilon_{2}$),
which were further interpolated (or extrapolated) while requiring that 
the finally resulting non-LTE abundance be consistent with the input 
theoretical abundance (cf. Section 4.2 in Takeda \& Takada-Hidai 1994). 

The values of the equivalent width ($W_{i}$), non-LTE abundance
($\log\epsilon_{i}^{\rm NLTE}$), and non-LTE correction ($\Delta_{i}$)
for each line $i$ of C, N, O, and Na, which were derived for all the 
122 spectra, are presented as the electronic data tables (the results 
for HD~?????? are given in the file named ``??????\_cnona.dat'') 
contained in the supplementary materials.

The final non-LTE abundance and the non-LTE correction for element X, 
$\log\epsilon$(X) and $\Delta_{\rm X}$ (X = C, N, O, and Na), 
were eventually obtained by averaging $\log\epsilon_{i}^{\rm NLTE}$ 
and $\Delta_{i}$ over each relevant line $i$. 
We also calculated the differential abundance relative to the Sun
as [X/H]~$\equiv \log\epsilon$(X)~$-$~$\log\epsilon_{\odot}$(X),
which we will mainly use in the discussion.
Regarding the reference solar abundances, we adopted the following 
values, which were derived from the solar flux spectra based on 
practically the same lines and the same atomic data by taking 
into account the non-LTE effect: 
$\log\epsilon_{\odot}$(C) = 8.51 (Takeda et al. 2013),
$\log\epsilon_{\odot}$(N) = 8.05 (newly derived for this study from 
the N~{\sc i} 8683 line with the non-LTE correction of $-0.05$~dex, 
for which the solar $W_{8683}$ was measured to be 6.1~m$\rm\AA$),
$\log\epsilon_{\odot}$(O) = 8.81 (Takeda \& Honda 2005), and
$\log\epsilon_{\odot}$(Na) = 6.32 (Takeda et al. 2003),
where the abundance values are expressed in the usual normalization
of $\log\epsilon$(H) = 12.00.
The resulting [C/H], [N/H], [O/H], and [Na/H] (along with
the corresponding $\Delta_{\rm C}$, $\Delta_{\rm N}$, $\Delta_{\rm O}$, 
and $\Delta_{\rm Na}$) derived for each of the 122 spectra are summarized 
in Table 2. As seen in this table, the extents of the (negative) non-LTE
corrections are not significant for C, O, and Na ($\la 0.1$~dex),
but rather considerable for N ($\sim$~0.2--0.4~dex).

\section{Discussion}

\subsection{Phase-Dependence of Physical Quantities}

The results of the physical parameters in the atmosphere ($V_{\rm rad}$, 
$T_{\rm eff}$, $\log g$, $\xi$, $v_{\rm M}$) as well as the elemental
abundances ([Fe/H], [C/H], [N/H], [O/H], and [Na/H]) are plotted
against the phase ($\phi$) in Fig. 7--18 for each of the program stars
(SU~Cas, SZ~Tau, RT~Aur, $\zeta$~Gem, FF~Aql, $\eta$~Aql, S~Sge, X~Cyg,
T~Vul, DT~Cyg, V1334~Cyg, and $\delta$~Cep).
Although we do not discuss these figures for each of the individual stars 
in detail, we can recognize several typical characteristics already known 
for Cepheid variables (see, e.g., Bersier, Burki \& Kurucz 1997), such as the
enhancement of $T_{\rm eff}$ and turbulent velocity fields ($\xi$ and $v_{\rm M}$
which are well-correlated) at the contraction or compressed phase of 
positive line-of-sight velocity (relative to the mean system velocity).

In order to check the results of our analysis, our values
of $T_{\rm eff}$, $\log g$, $\xi$, and [Fe/H] derived for $\delta$~Cep 
(one of the best studied representative Cepheids) and those of previous 
studies (Luck \& Lambert 1985; Kovtyukh, Komarov \& Depenchuk 1994; 
Fry \& Carney 1997; Kovtyukh \& Andrievsky 1999; 
Andrievsky, Luck \& Kovtyukh 2005) are overplotted against $\phi$ in Fig. 19, 
where we restricted the comparison to spectroscopically determined 
parameters derived in a similar manner to ours.
We can see several notable trends from this figure:\\
---  A remarkably good agreement with these literature values is
seen for $T_{\rm eff}$ (Fig. 19a) and [Fe/H] (Fig. 19d).\\
---  Our spectroscopic gravities are almost consistent with
Kovtyukh \& Andrievsky's (1999) ``non-standard'' results,
reflecting that both were derived in the same manner
(mainly based on Fe~{\sc ii} lines along with the limited use 
of only very weak Fe~{\sc i} lines ), while older results
using stronger Fe~{\sc i} lines (e.g., Fry \& Carney 1997) 
tend to be appreciably lower by $\sim 0.5$~dex (Fig. 19b). 
This confirms the argument of Kovtyukh \& Andrievsky (1999)
that the spectroscopic $\log g$ based on the new ``non-stadard'' 
approach are preferable, since those old spectroscopic 
$\log g$ values of Cepheids (conventionally determined) were 
often found to be significantly lower than the dynamical 
values determined from the empirically estimated mass and radius 
(see, e.g., Luck \& Lambert 1985).\\
--- Regarding the microturbulence, our $\xi$ values tend to be 
systematically higher by $\sim 1$~km~s$^{-1}$ in comparison with other 
literature results (Fig. 19c). The reason for this discrepancy
is not clear, since it is observed not only in comparison with older 
work (where Fe~{\sc i} lines were used to determine $\xi$) but also 
with the non-standard results of Kovtyukh \& Andrievsky (1999) 
(where $\xi$ was determined by Fe~{\sc ii} lines as in this study). 

\subsection{Correlation with Pulsation Period}

Given that our program stars were so chosen as to cover a wide range of 
pulsation periods ($P \sim$~2--16~days), it is worth examining
how our spectroscopically determined atmospheric parameters
depend on $P$, which is closely correlated with the absolute magnitude
(also with the stellar mass) through the period--luminosity relation.
Fig. 20 shows all the results of $T_{\rm eff}$, $\log g$, $\xi$, 
$v_{\rm M}$, and [Fe/H] derived for each of the 12 stars, plotted 
against $\log P$. 

We can see a clear trend of $T_{\rm eff}$ decreasing with an increase
in $\log P$ (Fig. 20a), which can be roughly represented by the 
linear relation
\begin{equation}
\log T_{\rm eff} = 3.84 - 0.10 \log P
\end{equation}
($T_{\rm eff}$ is in K and $P$ is in days)
with an uncertainty of a few hundred K.
This reflects the fact the Cepheid instability strip
is tilted in the theoretical $\log L$ vs. $\log T_{\rm eff}$ diagram,
in the sense that higher-$L$ Cepheids with longer $P$ tend to have 
lower $T_{\rm eff}$.\footnote{
Although Luck et al. (2008) concluded that the mean $T_{\rm eff}$ for 
s-Cepheids tends to be higher than that of classical Cepheids, 
such a trend can not be confirmed in Fig. 20a for the five s-Cepheids 
(SU~Cas, SZ~Tau, FF~Aql, DT~Cyg, and V1334~Cyg) included in our targets, 
which might be due to the incomplete phase coverage as well as the 
insufficient number of our program stars compared with their study
(9 s-Cepheids and 30 classical Cepheids).
}

We then discuss the $P$-dependence of $\log g$ (Fig. 20b).
The surface gravity is expressed in terms of $M$, $L$, and 
$T_{\rm eff}$ as
\begin{equation}
\log (g/g_{\odot}) = \log (M/M_{\odot}) - \log (L/L_{\odot}) 
+ 4\log (T_{\rm eff}/T_{\rm eff, \odot}).
\end{equation}
If we use Benedict et al.'s (2007) period--luminosity relation
for galactic Cepheids (cf. their fig. 5)
\begin{equation}
M_{V} = -1.62 -  2.43  \log P
\end{equation}
 we have
\begin{equation}
\log (L/L_{\odot}) = 2.54 -0.4 B.C. + 0.97 \log P,
\end{equation}
where $B.C.$ is the bolometric correction.
We then invoke Caputo et al.'s (2005) mass--luminosity relation 
for solar composition (cf. their eq. (2))
\begin{equation}
\log (L/L_{\odot}) = 0.72 + 3.35 \log (M/M_{\odot}).
\end{equation}
Combining Eq. (5) and (6), we get a mass--period relation as
\begin{equation}
\log (M/M_{\odot}) = 0.54 -0.12 B.C. + 0.29 \log P.
\end{equation}
Consequently, inserting  Eq. (2), (5), and (7) into Eq. (3),
we finally obtain
\begin{equation}
\log g = 2.75 + 0.28 B.C. -1.08 \log P \simeq 2.72 - 1.08 \log P
\end{equation}
as the $P$-dependence of dynamical $\log g$, where we assumed
$B.C. \simeq -0.1$ (actual values range from $\sim -0.2$ to $\sim 0.0$
depending on $T_{\rm eff}$, but insignificant; e.g., Flower 1996).
This relation is  also depicted in Fig. 20b. 
As seen from this figure, the spectroscopic $\log g$ values derived 
in this study tend to be systematically larger by this relation 
of dynamical $\log g$, which indicates that some of the relations 
we assumed above may not be appropriate or some systematic error 
may be involved in our $\log g$. 
Nevertheless, considering the large dispersion of observed 
$\log g$ along with the incomplete phase coverage of our data, 
we may regard that both are in tolerable consistency, despite 
the discrepancy especially in short-period Cepheids of 
$P\sim$~2--3~days.

Fig. 20c and 20d suggest that $\xi$ as well as $v_{\rm M}$ attain 
an apparent maximum at $\log P\sim 0.8$ and turn to decrease with
increasing $P$, which may contradict the intuitive expectation that
the atmospheric turbulence would progressively grow with a decrease 
of $\log g$ because of increased instability in lower-density condition.
We consider, however, that this should not be taken too seriously,
since the phase coverage of X~Cyg (having the longest $P$ of 16~days) is 
insufficient and our 9 spectra for this star correspond to the expanding 
and decelerating phase ($\phi$~0.2--0.4) of non-enhanced turbulence 
(Fig. 14).

The dispersion of [Fe/H] for each star in Fig. 20e suggests that
the precision of its determination for a spectrum at a given phase 
(as a by-product of atmospheric parameter determination based on 
Fe~{\sc i} and Fe~{\sc ii} lines) is on the order of $\sim 0.1$~dex,
though with rather large star-to-star differences (e.g., the reason 
for the appreciably larger [Fe/H] dispersion in S~Sge is because
our data correspond to $\phi \sim$~0.7--1.1, where the spectral 
variation is rapid and the analysis is comparatively more difficult).  
Overall, these [Fe/H] values scatter around $\sim 0$ without any 
systematic dependence of [Fe/H] upon $P$, which means that all the 
program stars have almost the solar metallicity irrespective of
the absolute magnitude and the mass.

Somewhat unexpectedly, such $P$-independent chemical homogeneity 
(i.e., small star-to-star abundance dispersion) was found also 
for the other elements. By averaging the [X/H]$_{i}$ results for 
each individual spectrum $i$, we computed the mean abundance 
$\langle$[X/H]$\rangle$ and the standard deviation $\sigma_{\rm X}$
for each of the 12 stars (X = C, N, O, Na, and Fe), which are given
in Table 2 (expressed in {\it italic} at the first line of 
each star's section). These $\langle$[C/H]$\rangle$, 
$\langle$[N/H]$\rangle$, $\langle$[O/H]$\rangle$, $\langle$[Na/H]$\rangle$, 
and $\langle$[Fe/H]$\rangle$ for each star are plotted against $\log P$ 
in Fig. 21a--e, where the extents of relevant $\pm \sigma$ are shown 
as error bars. These figures demonstrate that none of
these $\langle$[X/H]$\rangle$s show any significant $P$-dependence,
which are practically homogeneous within a dispersion of 
$\la$~0.1--0.2~dex. We may thus conclude 
$\langle$[C/H]$\rangle$~$\sim -0.3$, 
$\langle$[N/H]$\rangle$~$\sim$~+0.3--0.4, 
$\langle$[O/H]$\rangle$~$\sim 0$,  
$\langle$[Na/H]$\rangle$~$\sim +0.2$, and 
$\langle$[Fe/H]$\rangle$~$\sim 0$ 
as the abundance characteristics for these elements, 
which hold for all the program stars regardless of $\log P$.
We note that these values are almost consistent with  
the recent extensive studies for various Cepheid variables  
(Andrievsky et al. 2005; Luck \& Andrievsky 2004;
Kovtyukh et al. 2005a; Luck et al. 2008), where we can see that their 
C, N, O, and Na abundances are almost the $P$-independent with the
means (and the standard deviations) of  
$\langle$[C/H]$\rangle = -0.15$ ($\sigma = 0.08$; 38 stars excluding 
SV~Mon showing an exceptionally large deficiency of $-0.90$), 
$\langle$[N/H]$\rangle = +0.38$ ($\sigma = 0.17$; 6 stars), 
$\langle$[O/H]$\rangle = -0.08$ ($\sigma = 0.09$; 38 stars), and 
$\langle$[Na/H]$\rangle = +0.21$ ($\sigma = 0.07$; 39 stars). 

\subsection{Abundance Trends and Their Implications}

Now, since the abundances of key elements have been established,
we can discuss the main problem which motivated this investigation
(cf. Section 1): the nature of envelope mixing inferred from 
the characteristics of C, N, O, and Na abundances. 

According to the results derived in Section 5.2, C is mildly 
deficient by $\sim 0.3$~dex, N is appreciably enhanced by 
$\sim$~0.3--0.4~dex, Na is moderately overabundant by $\sim 0.2$~dex, 
while O (as well as Fe) is practically solar without any significant 
peculiarity (within an uncertainty of $\sim$~0.1~dex).
This tendency can be visually confirmed in Fig. 22a and Fig. 22b,
where $\langle$[N/H]$\rangle$ vs. $\langle$[C/H]$\rangle$
and $\langle$[Na/H]$\rangle$ vs. $\langle$[O/H]$\rangle$
correlations are plotted, respectively. 

Regarding the oxygen problem to be clarified in the first place,
we can confidently state that any significant mixing of ON-cycle 
product has not taken place in the envelope of Cepheid variables.
This means that the result of apparently subsolar [O/H] 
derived by Luck \& Lambert (1985) is most likely due to
the use of inappropriate atmospheric parameters (especially
too low $\log g$), as suggested by Kovtyukh \& Andrievsky (1999).
Then, the abundance peculiarities of C and N must have been 
caused mainly by the dredge-up of CN-cycle product.
The weak anti-correlation between the C and N abundances implied 
from Fig. 22a is qualitatively consistent with this scenario.

In order to check this quantitatively, the CNO abundances 
($\log\epsilon$) of our program stars are plotted 
in the $\langle\log\epsilon$(N)$\rangle$ 
vs. $\langle\log\epsilon$(C+O)$\rangle$ diagram 
and the $\langle\log\epsilon$(N)$\rangle$ vs. 
$\langle\log\epsilon$(C)$\rangle$ diagram  in Fig. 22c and Fig. 22d,
respectively,  where the expected relations under the condition 
that $\epsilon$(C+N+O) is conserved are also drawn.
Further, the resulting sums of $\log\epsilon$(C+N+O) evaluated
for the program stars are plotted against $\log P$ in Fig. 22e.
We can confirm that the observed abundances are almost 
on the expected curve in Fig. 22c (compare this figure with Fig. 1 of
Luck \& Lambert 1985), and that $\log\epsilon$(C+N+O) is
almost conserved at the solar value (Fig. 22f).

It should be remarked, however, that we can not rule out the possibility
of only a slight underabundance in O caused by the dredge-up of 
some (not significant) ON-cycle product. Actually, Fig. 22d implies
that $\log\epsilon$(N) vs. $\log\epsilon$(C) correlation can not 
necessarily be well described only by the conservation conditions
of $\log\epsilon$(C+N)~=~$\log\epsilon_{\odot}$(C+N) and
$\log\epsilon$(O)~=~$\log\epsilon_{\odot}$(O) for stars of
$\log\epsilon$(N)$\ga 8.5$. Rather, it is more reasonable to
consider that a mixing of some ON-cycled material could have 
caused a slight O-deficiency by several hundredths dex
(within $\la 0.1$~dex) in order to explain the 
$\log\epsilon$(N) values of these stars (Fig. 22d).

Finally, our results of moderate enhancement of Na by $\sim 0.2$~dex 
indicate that the NeNa-cycle product is 
dredged-up in the envelope of Cepheid variables.
Takeda \& Takada-Hidai (1994) derived [Na/H] = +0.15 ($\eta$~Aql)
and [Na/H] = $-0.07$ ($\zeta$~Gem) based on their non-LTE analysis of 
the Na~{\sc i} 8194.12 line.  Comparing these with the present results
of +0.16 ($\eta$~Aql) and +0.23 ($\zeta$~Gem), we can see an appreciable
discrepancy for the latter. However, their old analysis should be regarded 
as less reliable, where the atmospheric parameters corresponding to the 
observed phase were not properly determined but roughly assumed 
while consulting the literature values.

\subsection{Mass-Independent Mixing in Cepheids}

Before starting this investigation, we anticipated that a significantly 
large abundance scatter may be found at least for C and N (or possibly 
also for O), as reported in the pioneering work by Luck \& Lambert 
(1985; cf. Fig. 1 or Fig. 9 therein), and that such a diversity 
(if exists) might be correlated with $P$ (or equivalently, the stellar 
mass $M$). However, the considerably small star-to-star dispersion 
in the abundances of all elements (C, N, O, Na, Fe) was rather an 
unexpected result. Though, admittedly, a weak anti-correlation between 
N and C abundances (Fig. 22a, Fig. 22d) suggests that the extent of 
dredged-up material mixed in envelope may differ slightly from star 
to star, it does not appear to have anything to do with $M$, since 
[N/C] does not show any systematic dependence upon $P$ (Fig. 22e).
So, another important outcome of this study is the confirmation that 
the extent of mixing in the envelope of Cepheid variables does not 
depend upon the stellar mass.
Interestingly, essentially the same conclusion was reached by Kovtyukh, 
Wallerstein \& Andrievsky (2005b), who carried out an extensive spectroscopic 
abundance study for 16 distant Cepheids with a wide range of $P$  
($\sim$~3--27~days) and found that neither [C/N] ($\sim +0.6$) nor 
[Na/Fe] ($\sim +0.2$) show any systematic dependence upon $\log P$,
which is quite consistent with our results.

We should note, however, that this consequence is guaranteed
only for the pulsation variables within the Cepheid instability
strip, which should not be simply extended to non-variable 
supergiants or giants in different parameter ranges
(e.g., in terms of $T_{\rm eff}$ or $M$). 
For example, we can not say much about the argument of
Takeda \& Takada-Hidai (1994) (Na tends to be more enriched 
with an increase of mass for supergiants of 
$\sim$~10--25~$M_{\odot}$; cf. Fig. 7 therein), 
since the mass range of the program stars in this study 
is between $\sim 4 M_{\odot}$ and $\sim 8 M_{\odot}$
according to Eq. (7), 
and their sample was mostly non-variable supergiants.
Similarly, the systematic $\log g$-dependence of [Na/H] 
concluded by Andrievsky et al. (2002) was mainly due to
supergiants (not Cepheids) of $0.5 \la \log g \la 2$ 
(cf. their fig. 1), which is again less relevant to 
this study.
Besides, Smiljanic et al. (2006) concluded based on their
CNO abundance study for 19 non-variable supergiants 
covering a rather large range of stellar parameters
(4100~K $\la T_{\rm eff} \la$ 7500~K, 0.8 $\la \log g \la$ 2.5,
2 $\la M/M_{\odot} \la$ 13) that the [N/C] ratios tend to 
increase with $M$ and have considerably larger values 
(up to $\la +2$) than the theoretical expectation.
Further, Takeda, Sato \& Murata (2008) reported in their analysis
of 322 late-G and early-K giants
(4500~K $\la T_{\rm eff} \la$ 5500~K, 1.5 $\la \log g \la$ 3.5,
1 $\la M/M_{\odot} \la$ 5) that [C/Fe], [O/Fe], and 
[Na/Fe] show subsolar, subsolar, and supersolar tendency, 
respectively, with the degree of peculiarity increasing with $M$ 
(cf. Fig. 12 therein), which may imply that a mass-dependent 
dredge-up of CN-cycle, ON-cycle, and NeNa-cycle products 
may take place in the envelope of red giants.

Consequently, we would still consider it quite possible that 
the mixing-induced abundance peculiarities of non-variable 
intermediate-mass evolved stars in general are significantly 
diversified and may depend on $M$.
If so, we would have to find a reasonable answer to the question
``Why the absence of large star-to-star dispersion or of
$M$-dependence in the evolution-induced mixing is limited only 
to Cepheid variables?'', to which contributions by theoreticians 
are desirably awaited.

\section{Conclusion}

The mixing in the envelope of intermediate-mass F--G supergiants 
causing surface abundance peculiarities by the dredge-up
of H-burning products is not yet well understood. Especially, 
regarding oxygen, whether or not its anomaly exists due to 
non-canonical dredge-up of ON-cycled material is still controversial. 

In order to clarify this situation, we conducted an extensive 
spectroscopic study for selected 12 Cepheid variables of various 
pulsation periods ($\sim $~2--16~days), and determined the 
photospheric abundances of C, N, O, and Na, which are the key elements 
for investigating the dredge-up of H-burning products from the 
interior, based on 122 high-dispersion spectra ($\sim 10$ spectra
of different phases per target) of wide wavelength coverage
collected at Bohyunsan Astronomical Observatory. 

The atmospheric parameters ($T_{\rm eff}$, $\log g$, $\xi$, and
[Fe/H]) corresponding to each phase were determined spectroscopically 
from the equivalent widths of Fe~{\sc i} and Fe~{\sc ii} lines 
by the requirements of excitation equilibrium, ionization equilibrium,
and curve-of-growth matching. Practically, we applied Takeda et al.'s 
(2005) TGVIT program, while using only fairly weak Fe~{\sc i}lines 
($\la 30$~m$\rm\AA$) as suggested by Kovtyukh \& Andrievsky 
(1999) to avoid non-LTE-sensitive stronger Fe~{\sc i} lines.
The resulting parameters as well as the radial velocities were 
confirmed to show the well-known phase-dependent characteristics
of Cepheids.

The abundances of C, N, O, and Na were derived by applying 
the spectrum-synthesis fitting technique to three
wavelength regions (6143--6163~$\rm\AA$, 7110--7121~$\rm\AA$, 
and 8677--8697~$\rm\AA$) comprising 
C~{\sc i} 7111/7113/7115/7116/7119, O~{\sc i} 6155--8, 
N~{\sc i} 8680/8683/8686, and Na~{\sc i} 6154/6161 lines.
Then, from the equivalent widths inversely computed from
the fitting-based abundances, the final abundances were
obtained by taking into account the non-LTE effect.

The resulting abundances of these elements for 12 program stars 
turned out to show remarkably small star-to-star dispersions 
($\la$~0.1--0.2~dex) without any significant dependence upon the 
pulsation period: near-solar Fe ([Fe/H]~$\sim 0.0$), moderately 
underabundant C ([C/H]~$\sim -0.3$), appreciably overabundant N 
([N/H]~$\sim$~+0.4--0.5), and mildly supersolar Na ([Na/H]~$\sim +0.2$). 
Based on these results, we conclude as follows.\\
--- (1) These CNO abundance trends can be interpreted mainly as due to 
the canonical dredge-up of CN-cycled material, while the significant
non-canonical deep mixing of ON-cycled gas is ruled out (though only 
a slight mixing may still be possible).\\ 
--- (2) The mild but definite overabundance of Na suggests that 
the NeNa-cycle product is also dredged up.\\ 
--- (3) The extent of mixing-induced peculiarities in the envelope
of Cepheid variables is almost independent on $M_{V}$ 
as well as on $M$. However, given 
the observational suggestion that a significant diversity or a 
tendency of $M$-dependence exists in the C, N, O, and Na abundances 
of non-variable supergiants/giants of other types, the 
problem  of ``why such a homogeneity in the surface abundances 
is limited only to Cepheids'' is yet to be investigated further.

\section*{Acknowledgments}

I. Han acknowledges the financial support by KICOS through 
Korea--Ukraine joint research grant (grant  07-179).
B.-C. Lee acknowledges the Astrophysical Research Center for the Structure 
and Evolution of the Cosmos  (ARSEC, Sejong University) of the 
Korea Science and Engineering Foundation (KOSEF) through the Science
Research Center (SRC) program.

\setcounter{table}{0}
\begin{table*}
\begin{minipage}{130mm}
\small
\caption{Basic data of the program stars.}
\begin{center}
\begin{tabular}{cccccccccr}\hline
Name & HD\#  & $\alpha_{2000}$ & $\delta_{2000}$ & Cep.Type & Sp.Type & $V$ & Epoch & $P$ \\
(1) & (2) & (3) & (4) & (5) & (6) & (7) & (8) & (9) \\
\hline
SU Cas&017463&02:51:58.8 &+68:53:19 & s &F5Ib/II--F7Ib/II &  5.70--6.18  &50100.156 & 1.949319  \\
SZ Tau&029260&04:37:14.8 &+18:32:35 & s &F5Ib--F9.5Ib  &  6.33--6.75  &50101.605 & 3.14873     \\
RT Aur&045412&06:28:34.1 &+30:29:35 & cl &F4Ib--G1Ib    &  5.00--5.82  &50101.159 & 3.728115 \\
$\zeta$ Gem&052973&07:04:06.5 &+20:34:13 & cl &F7Ib--G3Ib  &  3.62--4.18  &50108.93~  &  10.15073   \\
FF Aql&176155&18:58:14.7 &+17:21:39 & s &F5Ia--F8Ia    &  5.18--5.68  &50102.387 &   4.470916   \\
$\eta$ Aql&187929&19:52:28.4 &+01:00:20 & cl &F6Ib--G4Ib   &  3.48--4.39  &50100.861 &  7.176641 \\
S Sge&188727&19:56:01.3 &+16:38:05 & cl &F6Ib--G5Ib    &  5.24--6.04  &50105.348 &  8.382086    \\
X Cyg&197572&20:43:24.2 &+35:35:16 & cl &F7Ib--G8Ib    &  5.85--6.91  &50106.014 &  16.386332    \\
T Vul&198726&20:51:28.2 &+28:15:02 & cl &F5Ib--G0Ib    &  5.41--6.09  &50101.410 &   4.435462    \\
DT Cyg&201078&21:06:30.2 &+31:11:05 & s &F5.5--F7Ib/II &  5.57--5.96  &50102.487 &   2.499215    \\
V1334 Cyg&203156&21:19:22.2 &+38:14:15 & s &F2Ib       &  5.77--5.96  &50102.549 &   3.332816    \\
$\delta$ Cep&213306&22:29:10.3 &+58:24:55 & cl &F5Ib--G1Ib &  3.48--4.37  &50102.860 &   5.366341  \\
\hline
\end{tabular}
\end{center}
Following the star name and HD number (serial number in the Henry--Draper catalog)
in Columns (1) and (2), the data of star coordinates (right ascension in HH:MM:SS.S 
and declination in deg:arcmin:arcsec), Cepheid type (``cl'' for classical Cepheids
and ``s'' for s-Cepheids), spectral type, and apparent visual magnitude (in mag) 
are presented in Columns (3)--(7), which were taken from the web site of 
General Catalogue of Variable Stars (http://www.sai.msu.su/gcvs/gcvs/index.htm). 
The last two Columns (8) and (9) give the epoch of brightness maximum expressed
in Julian day (JD~$-2400000$) and the pulsation period (in day), for which we consulted
Table 1 of Kiss (1998).
\end{minipage}
\end{table*}

\setcounter{table}{1}
\begin{table*}
\begin{minipage}{180mm}
\scriptsize
\caption{Results of stellar parameters and abundances derived from each spectrum of different phase.}
\begin{center}
\begin{tabular}{crcrccccr c@{ }c c@{ }c c@{ }c c@{ }c}\hline
Code & JD & $\phi$ & $V_{\rm rad}$ & $T_{\rm eff}$ & $\log g$ & [Fe] & 
$\xi$ & $v_{\rm M}$ & [C] & $\Delta_{\rm C}$ & [N] & $\Delta_{\rm N}$ &
[O] & $\Delta_{\rm O}$ & [Na] & $\Delta_{\rm Na}$ \\
(1) & (2) & (3) & (4) & (5) & (6) & (7) & (8) & (9) & (10) & (11) & (12) & (13) & (14) & (15) & (16) & (17) \\
\hline
\multicolumn{6}{l}{SU~Cas (HD~017463)} &\multicolumn{3}{l}{${\it +0.06 (\pm 0.09)}$} & \multicolumn{2}{l}{${\it -0.30 (\pm 0.05)}$} & \multicolumn{2}{l}{${\it +0.46 (\pm 0.05)}$} & \multicolumn{2}{l}{${\it +0.05 (\pm 0.06)}$} & \multicolumn{2}{l}{${\it +0.20 (\pm 0.03)}$} \\ 
017463-091002A&  6.99&  0.503&  $-$1.4& 6500& 3.43& +0.27& 3.6&  7.6& $-$0.26& $-$0.04& +0.55& $-$0.26& +0.11& $-$0.03& +0.28& $-$0.07 \\
017463-091002B&  7.22&  0.623&  +1.3& 6336& 3.04& +0.14& 3.7&  9.1& $-$0.27& $-$0.05& +0.48& $-$0.27& +0.09& $-$0.03& +0.21& $-$0.07 \\
017463-091002C&  7.30&  0.663&  +1.3& 6469& 3.46& +0.22& 4.2&  9.6& $-$0.23& $-$0.04& +0.55& $-$0.24& +0.15& $-$0.03& +0.25& $-$0.07 \\
017463-091003A&  7.98&  0.012& $-$14.8& 6833& 2.76& $-$0.08& 3.9& 10.9& $-$0.38& $-$0.07& +0.38& $-$0.41& $-$0.03& $-$0.05& +0.16& $-$0.07 \\
017463-091003B&  8.08&  0.062& $-$15.8& 6935& 2.88& $-$0.01& 3.8& 10.9& $-$0.37& $-$0.07& +0.36& $-$0.41& $-$0.07& $-$0.05& +0.20& $-$0.07 \\
017463-091003C&  8.16&  0.103& $-$14.9& 6853& 2.89& +0.00& 3.7& 10.4& $-$0.36& $-$0.07& +0.40& $-$0.41& $-$0.04& $-$0.05& +0.21& $-$0.07 \\
017463-091003D&  8.21&  0.132& $-$15.0& 6916& 3.19& +0.02& 4.1&  9.7& $-$0.31& $-$0.06& +0.43& $-$0.36& +0.01& $-$0.04& +0.23& $-$0.07 \\
017463-091003E&  8.26&  0.156& $-$14.0& 6632& 2.75& +0.00& 3.5&  9.6& $-$0.33& $-$0.07& +0.45& $-$0.39& +0.05& $-$0.05& +0.16& $-$0.07 \\
017463-091003F&  8.30&  0.178& $-$13.6& 6599& 2.75& +0.03& 3.5&  9.3& $-$0.33& $-$0.07& +0.45& $-$0.39& +0.03& $-$0.05& +0.17& $-$0.07 \\
017463-091003G&  8.34&  0.197& $-$12.7& 6660& 2.86& +0.04& 3.5&  9.0& $-$0.35& $-$0.07& +0.44& $-$0.38& +0.02& $-$0.05& +0.21& $-$0.07 \\
017463-091004A&  9.05&  0.563&  +0.3& 6279& 2.83& +0.08& 3.5&  7.5& $-$0.27& $-$0.06& +0.46& $-$0.30& +0.05& $-$0.04& +0.19& $-$0.07 \\
017463-091004B&  9.14&  0.605&  +1.7& 6279& 2.90& +0.09& 3.6&  8.0& $-$0.24& $-$0.06& +0.47& $-$0.28& +0.06& $-$0.04& +0.19& $-$0.07 \\
017463-091004C&  9.21&  0.643&  +2.6& 6278& 2.86& +0.08& 3.7&  8.5& $-$0.28& $-$0.06& +0.47& $-$0.29& +0.06& $-$0.04& +0.18& $-$0.07 \\
017463-091004D&  9.28&  0.677&  +2.6& 6410& 3.29& +0.17& 4.1&  9.0& $-$0.20& $-$0.04& +0.54& $-$0.25& +0.12& $-$0.03& +0.23& $-$0.07 \\
017463-091005A&  9.97&  0.034& $-$15.0& 6964& 3.14& +0.01& 4.1&  9.5& $-$0.32& $-$0.06& +0.42& $-$0.38& +0.01& $-$0.04& +0.19& $-$0.07 \\
017463-091005B& 10.22&  0.160& $-$14.6& 6596& 2.78& +0.00& 3.8&  8.8& $-$0.30& $-$0.07& +0.47& $-$0.38& +0.09& $-$0.05& +0.16& $-$0.07 \\
017463-091005C& 10.28&  0.190& $-$13.3& 6629& 2.90& +0.04& 3.6&  8.5& $-$0.30& $-$0.07& +0.47& $-$0.37& +0.08& $-$0.05& +0.19& $-$0.07 \\
\hline
\multicolumn{6}{l}{SZ~Tau (HD~029260)} &\multicolumn{3}{l}{${\it +0.05 (\pm 0.04)}$} & \multicolumn{2}{l}{${\it -0.26 (\pm 0.04)}$} & \multicolumn{2}{l}{${\it +0.48 (\pm 0.05)}$} & \multicolumn{2}{l}{${\it +0.03 (\pm 0.04)}$} & \multicolumn{2}{l}{${\it +0.22 (\pm 0.03)}$} \\ 
029260-091002A&  7.15&  0.704& +10.0& 6010& 2.42& +0.01& 4.3& 12.2& $-$0.24& $-$0.07& +0.54& $-$0.31& +0.07& $-$0.05& +0.18& $-$0.08 \\
029260-091002B&  7.24&  0.732&  +8.9& 6143& 2.71& +0.07& 4.9& 13.0& $-$0.22& $-$0.06& +0.57& $-$0.28& +0.11& $-$0.04& +0.22& $-$0.07 \\
029260-091002C&  7.32&  0.758&  +7.2& 6141& 2.42& $-$0.01& 4.5& 13.7& $-$0.29& $-$0.07& +0.46& $-$0.32& $-$0.01& $-$0.05& +0.18& $-$0.07 \\
029260-091003A&  8.17&  0.027& $-$11.0& 6405& 2.41& +0.06& 4.0& 12.1& $-$0.31& $-$0.08& +0.41& $-$0.38& $-$0.03& $-$0.06& +0.25& $-$0.07 \\
029260-091003B&  8.23&  0.045& $-$10.6& 6272& 2.24& +0.00& 3.9& 11.6& $-$0.29& $-$0.09& +0.46& $-$0.39& +0.01& $-$0.07& +0.21& $-$0.08 \\
029260-091003C&  8.27&  0.059&  $-$9.7& 6276& 2.35& +0.01& 4.0& 11.4& $-$0.25& $-$0.09& +0.48& $-$0.38& +0.04& $-$0.06& +0.21& $-$0.08 \\
029260-091003D&  8.32&  0.073&  $-$9.8& 6427& 2.81& +0.12& 4.5& 11.2& $-$0.17& $-$0.06& +0.53& $-$0.33& +0.07& $-$0.05& +0.28& $-$0.07 \\
029260-091004A&  9.15&  0.337&  $-$0.8& 6044& 2.50& +0.08& 3.9&  7.7& $-$0.25& $-$0.07& +0.45& $-$0.29& $-$0.04& $-$0.04& +0.25& $-$0.08 \\
029260-091004B&  9.22&  0.362&  +0.6& 5968& 2.46& +0.06& 3.8&  7.6& $-$0.20& $-$0.07& +0.47& $-$0.28& +0.02& $-$0.04& +0.23& $-$0.08 \\
029260-091004C&  9.29&  0.384&  +1.9& 5919& 2.30& +0.03& 3.8&  7.7& $-$0.26& $-$0.07& +0.47& $-$0.29& $-$0.01& $-$0.05& +0.21& $-$0.08 \\
029260-091005A& 10.15&  0.655& +11.7& 5961& 2.35& +0.06& 3.9& 11.1& $-$0.26& $-$0.07& +0.46& $-$0.30& +0.01& $-$0.05& +0.21& $-$0.08 \\
029260-091005B& 10.30&  0.704& +10.4& 6063& 2.36& +0.06& 4.3& 12.2& $-$0.32& $-$0.08& +0.42& $-$0.31& +0.05& $-$0.05& +0.21& $-$0.08 \\
\hline
\multicolumn{6}{l}{RT~Aur (HD~045412)} &\multicolumn{3}{l}{${\it +0.09 (\pm 0.07)}$} & \multicolumn{2}{l}{${\it -0.25 (\pm 0.07)}$} & \multicolumn{2}{l}{${\it +0.47 (\pm 0.05)}$} & \multicolumn{2}{l}{${\it -0.01 (\pm 0.05)}$} & \multicolumn{2}{l}{${\it +0.25 (\pm 0.05)}$} \\ 
045412-091002A&  7.18&  0.776& +37.2& 5695& 2.08& +0.01& 4.6& 10.5& $-$0.20& $-$0.07& +0.44& $-$0.25& +0.06& $-$0.04& +0.18& $-$0.07 \\
045412-091002B&  7.27&  0.800& +37.1& 5746& 2.18& +0.08& 4.6& 10.8& $-$0.23& $-$0.07& +0.41& $-$0.24& $-$0.01& $-$0.04& +0.20& $-$0.07 \\
045412-091003A&  8.18&  0.043&  +3.3& 7001& 2.81& +0.06& 4.5& 10.1& $-$0.34& $-$0.07& +0.45& $-$0.45& $-$0.08& $-$0.05& +0.23& $-$0.06 \\
045412-091003B&  8.24&  0.060&  +2.8& 7000& 2.85& +0.02& 5.0& 10.0& $-$0.35& $-$0.07& +0.45& $-$0.43& $-$0.06& $-$0.05& +0.22& $-$0.06 \\
045412-091003C&  8.29&  0.072&  +1.7& 6890& 2.61& +0.00& 4.4& 10.3& $-$0.35& $-$0.08& +0.46& $-$0.46& $-$0.05& $-$0.05& +0.22& $-$0.06 \\
045412-091003D&  8.33&  0.083&  +1.6& 7005& 3.03& +0.10& 4.8& 10.1& $-$0.27& $-$0.06& +0.51& $-$0.42& $-$0.03& $-$0.04& +0.24& $-$0.06 \\
045412-091004A&  9.16&  0.306& +12.3& 6269& 2.75& +0.17& 3.5&  6.7& $-$0.19& $-$0.07& +0.54& $-$0.32& +0.05& $-$0.04& +0.31& $-$0.08 \\
045412-091004B&  9.25&  0.329& +13.6& 6223& 2.69& +0.18& 3.5&  6.2& $-$0.20& $-$0.07& +0.51& $-$0.32& +0.01& $-$0.04& +0.30& $-$0.08 \\
045412-091004C&  9.32&  0.349& +15.0& 6250& 2.92& +0.22& 3.6&  5.8& $-$0.16& $-$0.06& +0.55& $-$0.29& +0.04& $-$0.04& +0.32& $-$0.07 \\
045412-091005A& 10.24&  0.595& +28.8& 5843& 2.43& +0.10& 4.0&  7.4& $-$0.21& $-$0.06& +0.42& $-$0.24& $-$0.01& $-$0.04& +0.24& $-$0.07 \\
\hline
\multicolumn{6}{l}{$\zeta$~Gem (HD~052973)} &\multicolumn{3}{l}{${\it +0.04 (\pm 0.03)}$} & \multicolumn{2}{l}{${\it -0.29 (\pm 0.10)}$} & \multicolumn{2}{l}{${\it +0.35 (\pm 0.05)}$} & \multicolumn{2}{l}{${\it -0.14 (\pm 0.11)}$} & \multicolumn{2}{l}{${\it +0.23 (\pm 0.03)}$} \\ 
052973-091002A&  7.20&  0.405& +17.5& 5321& 1.68& +0.05& 4.4&  8.2& $-$0.17& $-$0.07& +0.38& $-$0.19& $-$0.14& $-$0.03& +0.19& $-$0.07 \\
052973-091002B&  7.29&  0.413& +18.3& 5342& 1.90& +0.06& 4.4&  8.4& $-$0.12& $-$0.06& +0.40& $-$0.18& +0.09& $-$0.03& +0.19& $-$0.07 \\
052973-091003A&  8.25&  0.508& +22.6& 5452& 1.76& +0.02& 4.7& 11.2& $-$0.28& $-$0.07& +0.29& $-$0.20& $-$0.13& $-$0.04& +0.24& $-$0.07 \\
052973-091003B&  8.29&  0.512& +22.6& 5462& 1.72& +0.04& 4.6& 11.3& $-$0.29& $-$0.08& +0.29& $-$0.21& $-$0.14& $-$0.04& +0.25& $-$0.07 \\
052973-091003C&  8.33&  0.516& +22.5& 5501& 1.78& +0.06& 4.5& 11.4& $-$0.27& $-$0.08& +0.28& $-$0.22& $-$0.22& $-$0.04& +0.26& $-$0.07 \\
052973-091004A&  9.25&  0.607& +17.5& 5691& 2.08& +0.09& 5.1& 11.9& $-$0.33& $-$0.07& +0.38& $-$0.23& $-$0.20& $-$0.04& +0.27& $-$0.07 \\
052973-091004B&  9.33&  0.615& +16.9& 5581& 1.62& $-$0.01& 4.6& 11.8& $-$0.41& $-$0.09& +0.35& $-$0.26& $-$0.32& $-$0.05& +0.22& $-$0.07 \\
052973-091005A& 10.25&  0.706&  +5.8& 5688& 1.79& +0.00& 4.8&  9.8& $-$0.41& $-$0.09& +0.43& $-$0.28& $-$0.07& $-$0.05& +0.21& $-$0.07 \\
\hline
\multicolumn{6}{l}{FF~Aql (HD~176155)} &\multicolumn{3}{l}{${\it +0.04 (\pm 0.06)}$} & \multicolumn{2}{l}{${\it -0.34 (\pm 0.03)}$} & \multicolumn{2}{l}{${\it +0.54 (\pm 0.02)}$} & \multicolumn{2}{l}{${\it -0.06 (\pm 0.02)}$} & \multicolumn{2}{l}{${\it +0.27 (\pm 0.02)}$} \\ 
176155-091002A&  6.93&  0.356& $-$18.8& 6246& 2.57& +0.16& 5.1&  8.0& $-$0.27& $-$0.07& +0.58& $-$0.33& $-$0.02& $-$0.05& +0.31& $-$0.07 \\
176155-091003A&  7.92&  0.578&  $-$9.7& 6081& 2.28& +0.02& 5.5&  9.1& $-$0.35& $-$0.08& +0.52& $-$0.32& $-$0.07& $-$0.05& +0.25& $-$0.07 \\
176155-091003B&  7.99&  0.593&  $-$9.7& 6099& 2.29& +0.03& 5.5&  9.2& $-$0.34& $-$0.08& +0.51& $-$0.32& $-$0.09& $-$0.05& +0.26& $-$0.07 \\
176155-091004A&  8.91&  0.799& $-$13.8& 6198& 2.13& $-$0.02& 5.7&  9.8& $-$0.35& $-$0.09& +0.55& $-$0.38& $-$0.06& $-$0.06& +0.25& $-$0.07 \\
176155-091004B&  8.96&  0.810& $-$13.9& 6256& 2.16& +0.01& 5.5&  9.8& $-$0.36& $-$0.09& +0.54& $-$0.39& $-$0.10& $-$0.06& +0.27& $-$0.07 \\
176155-091004C&  9.07&  0.834& $-$16.5& 6294& 2.20& +0.00& 5.6&  9.7& $-$0.36& $-$0.09& +0.55& $-$0.39& $-$0.06& $-$0.06& +0.27& $-$0.07 \\
176155-091005A&  9.93&  0.026& $-$25.6& 6612& 2.54& +0.06& 5.7&  9.4& $-$0.33& $-$0.07& +0.56& $-$0.40& $-$0.04& $-$0.05& +0.30& $-$0.06 \\
\hline
\multicolumn{6}{l}{$\eta$~Aql (HD~187929)} &\multicolumn{3}{l}{${\it +0.09 (\pm 0.12)}$} & \multicolumn{2}{l}{${\it -0.20 (\pm 0.08)}$} & \multicolumn{2}{l}{${\it +0.31 (\pm 0.06)}$} & \multicolumn{2}{l}{${\it -0.10 (\pm 0.14)}$} & \multicolumn{2}{l}{${\it +0.16 (\pm 0.07)}$} \\ 
187929-091002A&  6.95&  0.553& $-$11.4& 5591& 2.05& +0.14& 4.3& 10.2& $-$0.16& $-$0.07& +0.31& $-$0.22& +0.07& $-$0.04& +0.17& $-$0.07 \\
187929-091003A&  7.94&  0.691&  +3.0& 5735& 2.55& +0.26& 4.8& 11.7& $-$0.07& $-$0.05& +0.24& $-$0.18& +0.04& $-$0.03& +0.29& $-$0.07 \\
187929-091003B&  8.00&  0.700&  +4.3& 5542& 2.01& +0.15& 4.5& 12.0& $-$0.14& $-$0.07& +0.23& $-$0.20& $-$0.05& $-$0.04& +0.19& $-$0.07 \\
187929-091004A&  8.92&  0.828&  +8.9& 5649& 1.67& $-$0.03& 5.1& 16.5& $-$0.27& $-$0.09& +0.29& $-$0.26& $-$0.28& $-$0.05& +0.10& $-$0.07 \\
187929-091004B&  8.97&  0.835&  +7.8& 5799& 2.22& +0.07& 5.7& 16.9& $-$0.19& $-$0.07& +0.33& $-$0.23& $-$0.15& $-$0.04& +0.17& $-$0.07 \\
187929-091004C&  9.08&  0.850&  +5.1& 5654& 1.45& $-$0.14& 5.0& 16.6& $-$0.32& $-$0.11& +0.31& $-$0.29& $-$0.32& $-$0.05& +0.04& $-$0.07 \\
187929-091005A&  9.94&  0.970& $-$29.4& 6502& 2.61& +0.14& 5.4& 12.5& $-$0.24& $-$0.07& +0.44& $-$0.34& +0.02& $-$0.05& +0.18& $-$0.07 \\
\hline
\end{tabular}
\end{center}
\end{minipage}
\end{table*}

\setcounter{table}{1}
\begin{table*}
\begin{minipage}{180mm}
\scriptsize
\caption{(Continued.)}
\begin{center}
\begin{tabular}{crcrccccr c@{ }c c@{ }c c@{ }c c@{ }c}\hline
Code & JD & $\phi$ & $V_{\rm rad}$ & $T_{\rm eff}$ & $\log g$ & [Fe] & 
$\xi$ & $v_{\rm M}$ & [C] & $\Delta_{\rm C}$ & [N] & $\Delta_{\rm N}$ &
[O] & $\Delta_{\rm O}$ & [Na] & $\Delta_{\rm Na}$ \\
(1) & (2) & (3) & (4) & (5) & (6) & (7) & (8) & (9) & (10) & (11) & (12) & (13) & (14) & (15) & (16) & (17) \\
\hline
\multicolumn{6}{l}{S~Sge (HD~188727)} &\multicolumn{3}{l}{${\it +0.02 (\pm 0.03)}$} & \multicolumn{2}{l}{${\it -0.25 (\pm 0.03)}$} & \multicolumn{2}{l}{${\it +0.39 (\pm 0.10)}$} & \multicolumn{2}{l}{${\it -0.10 (\pm 0.16)}$} & \multicolumn{2}{l}{${\it +0.15 (\pm 0.02)}$} \\ 
188727-091002A&  6.95&  0.702& +22.3& 5410& 1.59& $-$0.01& 4.2& 11.6& $-$0.26& $-$0.08& +0.21& $-$0.20& $-$0.24& $-$0.04& +0.14& $-$0.07 \\
188727-091003A&  7.94&  0.820& +20.0& 5750& 1.83& +0.04& 5.2& 14.0& $-$0.24& $-$0.09& +0.28& $-$0.26& $-$0.27& $-$0.05& +0.12& $-$0.07 \\
188727-091003B&  8.01&  0.827& +18.4& 5690& 1.66& $-$0.06& 5.5& 14.0& $-$0.24& $-$0.10& +0.33& $-$0.27& $-$0.22& $-$0.05& +0.10& $-$0.07 \\
188727-091003C&  8.10&  0.839& +15.3& 5878& 1.93& +0.04& 5.6& 13.7& $-$0.31& $-$0.09& +0.28& $-$0.28& $-$0.34& $-$0.05& +0.17& $-$0.07 \\
188727-091004A&  8.94&  0.939&  $-$9.3& 6176& 2.12& +0.01& 5.4& 10.8& $-$0.26& $-$0.09& +0.44& $-$0.36& +0.02& $-$0.07& +0.15& $-$0.07 \\
188727-091004B&  8.97&  0.943& $-$10.4& 6175& 2.15& +0.04& 5.1& 10.8& $-$0.23& $-$0.09& +0.49& $-$0.36& +0.03& $-$0.07& +0.15& $-$0.07 \\
188727-091004C&  9.08&  0.956& $-$12.1& 6238& 2.19& +0.04& 5.0& 10.7& $-$0.24& $-$0.09& +0.49& $-$0.37& +0.04& $-$0.07& +0.16& $-$0.07 \\
188727-091005A&  9.94&  0.059& $-$12.9& 6109& 1.98& +0.02& 5.0&  9.7& $-$0.21& $-$0.10& +0.49& $-$0.38& +0.07& $-$0.08& +0.14& $-$0.08 \\
188727-091005B& 10.07&  0.074& $-$12.2& 6136& 2.08& +0.06& 4.8&  9.5& $-$0.21& $-$0.10& +0.48& $-$0.37& +0.05& $-$0.07& +0.17& $-$0.07 \\
\hline
\multicolumn{6}{l}{X~Cyg (HD~197572)} &\multicolumn{3}{l}{${\it +0.09 (\pm 0.07)}$} & \multicolumn{2}{l}{${\it -0.30 (\pm 0.04)}$} & \multicolumn{2}{l}{${\it +0.49 (\pm 0.05)}$} & \multicolumn{2}{l}{${\it +0.01 (\pm 0.06)}$} & \multicolumn{2}{l}{${\it +0.21 (\pm 0.04)}$} \\ 
197572-091002A&  7.12&  0.200& $-$10.0& 5537& 1.67& +0.14& 4.2&  9.5& $-$0.34& $-$0.08& +0.59& $-$0.29& +0.08& $-$0.06& +0.21& $-$0.07 \\
197572-091003a&  7.95&  0.250&  $-$4.2& 5349& 1.44& +0.09& 4.0&  8.3& $-$0.33& $-$0.09& +0.53& $-$0.24& +0.11& $-$0.05& +0.19& $-$0.07 \\
197572-091003b&  8.02&  0.254&  $-$4.3& 5331& 1.32& +0.06& 3.9&  8.2& $-$0.34& $-$0.09& +0.52& $-$0.25& +0.05& $-$0.05& +0.19& $-$0.07 \\
197572-091003c&  8.11&  0.260&  $-$3.5& 5509& 1.72& +0.18& 4.0&  8.0& $-$0.34& $-$0.08& +0.44& $-$0.25& $-$0.07& $-$0.05& +0.28& $-$0.07 \\
197572-091004A&  8.99&  0.314&  +0.6& 5059& 0.81& $-$0.07& 3.7&  7.3& $-$0.32& $-$0.10& +0.47& $-$0.22& +0.01& $-$0.05& +0.13& $-$0.05 \\
197572-091004B&  9.09&  0.320&  +1.4& 5167& 1.19& +0.05& 3.8&  7.2& $-$0.28& $-$0.09& +0.50& $-$0.21& +0.01& $-$0.04& +0.17& $-$0.06 \\
197572-091004C&  9.17&  0.325&  +2.2& 5243& 1.33& +0.09& 3.8&  7.0& $-$0.29& $-$0.09& +0.48& $-$0.22& $-$0.03& $-$0.04& +0.20& $-$0.07 \\
197572-091005A& 10.00&  0.375&  +6.7& 5265& 1.54& +0.14& 3.8&  6.8& $-$0.24& $-$0.08& +0.44& $-$0.20& $-$0.06& $-$0.04& +0.26& $-$0.07 \\
197572-091005B& 10.09&  0.381&  +7.0& 5254& 1.56& +0.14& 3.9&  6.8& $-$0.22& $-$0.08& +0.41& $-$0.19& +0.03& $-$0.04& +0.26& $-$0.07 \\
\hline
\multicolumn{6}{l}{T~Vul (HD~198726)} &\multicolumn{3}{l}{${\it -0.03 (\pm 0.03)}$} & \multicolumn{2}{l}{${\it -0.29 (\pm 0.05)}$} & \multicolumn{2}{l}{${\it +0.35 (\pm 0.07)}$} & \multicolumn{2}{l}{${\it -0.08 (\pm 0.07)}$} & \multicolumn{2}{l}{${\it +0.09 (\pm 0.02)}$} \\ 
198726-091002A&  6.96&  0.530&  +9.9& 5821& 2.46& +0.04& 4.5& 10.0& $-$0.22& $-$0.06& +0.34& $-$0.22& $-$0.06& $-$0.03& +0.10& $-$0.07 \\
198726-091003A&  7.96&  0.754& +16.8& 5854& 2.23& $-$0.03& 5.2& 12.5& $-$0.29& $-$0.07& +0.24& $-$0.24& $-$0.15& $-$0.04& +0.12& $-$0.07 \\
198726-091003B&  8.03&  0.770& +16.6& 5847& 2.25& $-$0.06& 5.4& 12.6& $-$0.24& $-$0.07& +0.28& $-$0.24& $-$0.10& $-$0.04& +0.09& $-$0.07 \\
198726-091003C&  8.12&  0.792& +15.0& 5917& 2.20& $-$0.03& 4.9& 12.7& $-$0.33& $-$0.08& +0.27& $-$0.27& $-$0.23& $-$0.04& +0.09& $-$0.07 \\
198726-091004A&  8.95&  0.979& $-$17.5& 6558& 2.39& $-$0.09& 4.7& 11.6& $-$0.35& $-$0.08& +0.37& $-$0.39& $-$0.04& $-$0.06& +0.07& $-$0.07 \\
198726-091004B&  9.00&  0.990& $-$17.6& 6605& 2.48& $-$0.05& 4.4& 11.5& $-$0.33& $-$0.08& +0.38& $-$0.40& $-$0.07& $-$0.06& +0.08& $-$0.07 \\
198726-091004C&  9.10&  0.013& $-$18.3& 6562& 2.48& $-$0.04& 4.6& 11.2& $-$0.33& $-$0.08& +0.39& $-$0.39& $-$0.04& $-$0.06& +0.10& $-$0.07 \\
198726-091005A&  9.98&  0.211&  $-$7.5& 6036& 2.38& $-$0.02& 4.3&  8.9& $-$0.22& $-$0.07& +0.46& $-$0.30& +0.02& $-$0.05& +0.07& $-$0.08 \\
198726-091005B& 10.10&  0.237&  $-$6.8& 6010& 2.31& $-$0.02& 4.3&  8.7& $-$0.28& $-$0.08& +0.39& $-$0.30& $-$0.02& $-$0.05& +0.06& $-$0.08 \\
\hline
\multicolumn{6}{l}{DT~Cyg (HD~201078)} &\multicolumn{3}{l}{${\it +0.08 (\pm 0.06)}$} & \multicolumn{2}{l}{${\it -0.19 (\pm 0.05)}$} & \multicolumn{2}{l}{${\it +0.49 (\pm 0.03)}$} & \multicolumn{2}{l}{${\it +0.05 (\pm 0.02)}$} & \multicolumn{2}{l}{${\it +0.25 (\pm 0.03)}$} \\ 
201078-091003A&  7.97&  0.820&  +2.6& 6245& 2.43& $-$0.02& 4.1& 10.1& $-$0.24& $-$0.08& +0.44& $-$0.34& +0.04& $-$0.06& +0.18& $-$0.07 \\
201078-091003B&  8.04&  0.850&  +1.0& 6341& 2.58& $-$0.01& 4.3& 10.4& $-$0.24& $-$0.07& +0.44& $-$0.34& +0.04& $-$0.05& +0.23& $-$0.07 \\
201078-091003C&  8.13&  0.886&  $-$0.5& 6448& 2.82& +0.10& 4.2& 10.7& $-$0.17& $-$0.07& +0.49& $-$0.33& +0.03& $-$0.05& +0.26& $-$0.07 \\
201078-091003D&  8.19&  0.911&  $-$2.0& 6502& 2.92& +0.14& 4.5& 11.1& $-$0.08& $-$0.06& +0.47& $-$0.32& +0.07& $-$0.05& +0.28& $-$0.07 \\
201078-091004A&  9.01&  0.239&  $-$4.7& 6386& 2.75& +0.11& 3.9&  8.3& $-$0.19& $-$0.07& +0.51& $-$0.34& +0.07& $-$0.05& +0.27& $-$0.07 \\
201078-091004B&  9.11&  0.278&  $-$3.5& 6292& 2.65& +0.08& 3.8&  8.1& $-$0.19& $-$0.07& +0.52& $-$0.34& +0.08& $-$0.05& +0.25& $-$0.07 \\
201078-091004C&  9.18&  0.307&  $-$3.0& 6421& 3.04& +0.18& 4.0&  7.8& $-$0.15& $-$0.06& +0.55& $-$0.30& +0.08& $-$0.04& +0.30& $-$0.07 \\
201078-091005A& 10.02&  0.642&  +5.2& 6181& 2.71& +0.07& 4.1&  8.6& $-$0.20& $-$0.06& +0.49& $-$0.29& +0.03& $-$0.04& +0.24& $-$0.07 \\
201078-091005B& 10.11&  0.680&  +5.0& 6155& 2.52& +0.04& 4.0&  9.1& $-$0.26& $-$0.07& +0.51& $-$0.32& +0.02& $-$0.05& +0.21& $-$0.07 \\
\hline
\multicolumn{6}{l}{V1334~Cyg (HD~203156)} &\multicolumn{3}{l}{${\it +0.01 (\pm 0.06)}$} & \multicolumn{2}{l}{${\it -0.24 (\pm 0.03)}$} & \multicolumn{2}{l}{${\it +0.44 (\pm 0.03)}$} & \multicolumn{2}{l}{${\it +0.03 (\pm 0.03)}$} & \multicolumn{2}{l}{${\it +0.11 (\pm 0.04)}$} \\ 
203156-091003A&  8.06&  0.886& $-$22.4& 6286& 2.20& $-$0.06& 4.0& 14.7& $-$0.26& $-$0.10& +0.48& $-$0.40& +0.05& $-$0.07& +0.03& $-$0.07 \\
203156-091003B&  8.14&  0.911& $-$22.6& 6496& 2.60& +0.03& 4.3& 14.7& $-$0.26& $-$0.08& +0.46& $-$0.38& +0.04& $-$0.06& +0.13& $-$0.07 \\
203156-091003C&  8.20&  0.927& $-$22.6& 6503& 2.65& +0.05& 4.4& 14.6& $-$0.28& $-$0.07& +0.45& $-$0.37& +0.07& $-$0.06& +0.12& $-$0.07 \\
203156-091004A&  8.93&  0.145& $-$19.0& 6260& 2.33& $-$0.03& 4.0& 14.4& $-$0.23& $-$0.09& +0.43& $-$0.36& +0.04& $-$0.06& +0.09& $-$0.07 \\
203156-091004B&  9.03&  0.176& $-$19.2& 6231& 2.21& $-$0.01& 3.7& 14.5& $-$0.25& $-$0.10& +0.42& $-$0.38& +0.00& $-$0.07& +0.09& $-$0.08 \\
203156-091004C&  9.12&  0.204& $-$17.9& 6388& 2.66& +0.07& 4.1& 14.6& $-$0.20& $-$0.07& +0.45& $-$0.34& +0.02& $-$0.05& +0.15& $-$0.07 \\
203156-091004D&  9.19&  0.226& $-$17.1& 6246& 2.30& $-$0.03& 4.0& 14.4& $-$0.27& $-$0.09& +0.43& $-$0.37& +0.01& $-$0.06& +0.10& $-$0.08 \\
203156-091005A& 10.04&  0.479& $-$12.2& 6249& 2.26& $-$0.04& 3.9& 14.8& $-$0.22& $-$0.10& +0.39& $-$0.37& $-$0.03& $-$0.06& +0.08& $-$0.08 \\
203156-091005C& 10.18&  0.523& $-$11.5& 6499& 2.91& +0.11& 4.5& 15.0& $-$0.18& $-$0.06& +0.47& $-$0.33& +0.04& $-$0.05& +0.18& $-$0.07 \\
\hline
\multicolumn{6}{l}{$\delta$~Cep (HD~213306)} &\multicolumn{3}{l}{${\it +0.04 (\pm 0.05)}$} & \multicolumn{2}{l}{${\it -0.22 (\pm 0.05)}$} & \multicolumn{2}{l}{${\it +0.39 (\pm 0.07)}$} & \multicolumn{2}{l}{${\it +0.02 (\pm 0.06)}$} & \multicolumn{2}{l}{${\it +0.17 (\pm 0.04)}$} \\ 
213306-091002A&  6.98&  0.502& $-$14.1& 5706& 2.19& +0.07& 3.9&  7.0& $-$0.23& $-$0.07& +0.40& $-$0.24& $-$0.01& $-$0.04& +0.15& $-$0.07 \\
213306-091002B&  7.22&  0.545& $-$11.5& 5713& 2.31& +0.09& 4.2&  7.6& $-$0.18& $-$0.06& +0.41& $-$0.22& +0.04& $-$0.04& +0.17& $-$0.07 \\
213306-091003A&  7.98&  0.687&  $-$4.1& 5682& 2.06& +0.11& 4.5&  9.8& $-$0.24& $-$0.07& +0.29& $-$0.23& $-$0.03& $-$0.04& +0.21& $-$0.07 \\
213306-091003B&  8.07&  0.705&  $-$3.8& 5661& 2.12& +0.10& 4.6& 10.2& $-$0.20& $-$0.07& +0.33& $-$0.22& +0.02& $-$0.04& +0.19& $-$0.07 \\
213306-091003C&  8.15&  0.720&  $-$1.9& 5666& 2.14& +0.09& 4.8& 10.5& $-$0.17& $-$0.07& +0.34& $-$0.22& +0.03& $-$0.04& +0.18& $-$0.07 \\
213306-091003D&  8.21&  0.730&  $-$1.5& 5625& 1.98& +0.01& 4.8& 10.8& $-$0.18& $-$0.07& +0.33& $-$0.23& +0.04& $-$0.04& +0.17& $-$0.07 \\
213306-091003E&  8.26&  0.739&  $-$1.6& 5633& 1.92& +0.03& 4.8& 11.0& $-$0.17& $-$0.08& +0.30& $-$0.23& +0.01& $-$0.04& +0.17& $-$0.07 \\
213306-091003F&  8.30&  0.747&  $-$0.1& 5658& 1.99& +0.03& 4.8& 11.2& $-$0.14& $-$0.08& +0.32& $-$0.23& +0.01& $-$0.04& +0.18& $-$0.07 \\
213306-091004A&  8.96&  0.871& $-$11.2& 6330& 2.66& +0.10& 6.0& 11.7& $-$0.28& $-$0.06& +0.32& $-$0.28& $-$0.09& $-$0.04& +0.24& $-$0.07 \\
213306-091004B&  9.05&  0.886& $-$17.7& 6319& 2.52& +0.03& 5.8& 10.9& $-$0.27& $-$0.07& +0.40& $-$0.31& $-$0.01& $-$0.05& +0.20& $-$0.07 \\
213306-091004C&  9.13&  0.902& $-$22.7& 6323& 2.50& +0.01& 5.6& 10.4& $-$0.22& $-$0.07& +0.48& $-$0.33& +0.07& $-$0.05& +0.16& $-$0.07 \\
213306-091004D&  9.21&  0.916& $-$27.1& 6545& 2.65& +0.05& 5.6& 10.3& $-$0.26& $-$0.07& +0.43& $-$0.34& +0.02& $-$0.05& +0.20& $-$0.07 \\
213306-091004E&  9.27&  0.929& $-$31.1& 6663& 2.51& $-$0.03& 5.3& 10.4& $-$0.33& $-$0.07& +0.38& $-$0.38& $-$0.07& $-$0.05& +0.18& $-$0.06 \\
213306-091005A&  9.96&  0.058& $-$36.9& 6515& 2.31& $-$0.09& 5.0&  9.3& $-$0.27& $-$0.09& +0.44& $-$0.41& $-$0.02& $-$0.06& +0.08& $-$0.07 \\
213306-091005B& 10.21&  0.104& $-$34.7& 6427& 2.44& +0.08& 4.5&  9.4& $-$0.22& $-$0.08& +0.47& $-$0.38& +0.07& $-$0.06& +0.19& $-$0.07 \\
213306-091005C& 10.27&  0.114& $-$34.3& 6189& 2.12& +0.05& 4.4&  9.2& $-$0.21& $-$0.10& +0.56& $-$0.40& +0.18& $-$0.08& +0.10& $-$0.08 \\
\hline
\end{tabular}
\end{center}
Column (1) --- spectrum code (??????-yymmdd\#), where ?????? is the HD number, yymmdd is the observation
date (UT), and \# denotes the spectrum turn of a star for the day (A $\cdots$ 1st, B $\cdots$ 2nd, etc.).
Column (2) --- observational time in the Julian day expressed as JD~$-$~24455100. 
Column (3) --- pulsation phase.
Column (4) --- heliocentric radial velocity (in km~s$^{-1}$). In Columns (5)--(9) are given 
the results of atmospheric parameters, $T_{\rm eff}$ (effective temperature, in K), 
$\log g$ (surface gravity, in cm~s$^{-2}$), [Fe/H] (logarithmic Fe abundance relative to the Sun), 
$\xi$ (microturbulence, in km~s$^{-1}$), and $v_{\rm M}$ (macrobroadening velocity in km~s$^{-1}$ 
derived from 6143--6163~$\rm\AA$ fitting), respectively.    
The last 8 Columns present the final results of [X/H] (logarithmic abundance of element X 
relative to the Sun) and $\Delta_{\rm X}$ (non-LTE correction) for C (10, 11), N (12, 13), 
O (14, 15), and Na (16, 17). The mean abundance ($\langle$[X/H]$\rangle$) averaged over each 
of the phases along with the standard deviation ($\sigma_{\rm X}$) are also given at the first line of 
each section (expressed in $italic$).
\end{minipage}
\end{table*}

\setcounter{table}{2}
\begin{table*}
\begin{minipage}{160mm}
\small
\caption{Adopted atomic data of important spectral lines.}
\begin{center}
\begin{tabular}{ccrccccl}\hline\hline
Species & $\lambda$ ($\rm\AA$) & $\chi_{\rm low}$ (eV)& $\log gf$ & Gammar & Gammas & Gammaw & Remark\\
(1) & (2) & (3) & (4) & (5) & (6) & (7) & (8) \\
\hline
\multicolumn{8}{c}{[6143--6163~$\rm\AA$ fitting]} \\
 Si~{\sc i}  &6145.016 &  5.616 & $-$0.820 & (7.77)&($-$4.45)&($-$7.05) & \\
 Fe~{\sc ii} &6147.741 &  3.889 & $-$2.721 &  8.53 & $-$6.53 & $-$7.88  & \\
 Fe~{\sc ii} &6149.258 &  3.889 & $-$2.724 &  8.53 & $-$6.53 & $-$7.88  & \\
 Fe~{\sc ii} &6150.098 &  3.221 & $-$4.754 &  8.54 & $-$6.54 & $-$7.91  & \\
 Fe~{\sc i}  &6151.617 &  2.176 & $-$3.299 &  8.19 & $-$6.20 & $-$7.82  & \\
 Na~{\sc i}  &6154.226 &  2.102 & $-$1.560 &  7.85 & $-$4.39 &($-$7.29) & Na~{\sc i} 6154\\
 Si~{\sc i}  &6155.134 &  5.619 & $-$0.400 & (7.77)&($-$4.45)&($-$7.05) & \\
 Si~{\sc i}  &6155.693 &  5.619 & $-$1.690 & (7.77)&($-$4.45)&($-$7.05) & \\
 O~{\sc i}   &6155.961 & 10.740 & $-$1.401 &  7.60 & $-$3.96 &($-$7.23) & O~{\sc i} 6155--8 \\
 O~{\sc i}   &6155.971 & 10.740 & $-$1.051 &  7.61 & $-$3.96 &($-$7.23) & O~{\sc i} 6155--8 \\
 O~{\sc i}   &6155.989 & 10.740 & $-$1.161 &  7.61 & $-$3.96 &($-$7.23) & O~{\sc i} 6155--8 \\
 Ca~{\sc i}  &6156.023 &  2.521 & $-$2.200 &  7.49 & $-$4.69 & $-$7.50  & \\
 O~{\sc i}   &6156.737 & 10.740 & $-$1.521 &  7.61 & $-$3.96 &($-$7.23) & O~{\sc i} 6155--8 \\
 O~{\sc i}   &6156.755 & 10.740 & $-$0.931 &  7.61 & $-$3.96 &($-$7.23) & O~{\sc i} 6155--8 \\
 O~{\sc i}   &6156.778 & 10.740 & $-$0.731 &  7.62 & $-$3.96 &($-$7.23) & O~{\sc i} 6155--8 \\
 Fe~{\sc i}  &6157.725 &  4.076 & $-$1.260 &  7.70 & $-$6.06 & $-$7.84  & \\
 O~{\sc i}   &6158.149 & 10.741 & $-$1.891 &  7.62 & $-$3.96 &($-$7.23) & O~{\sc i} 6155--8 \\
 O~{\sc i}   &6158.172 & 10.741 & $-$1.031 &  7.62 & $-$3.96 &($-$7.23) & O~{\sc i} 6155--8 \\
 O~{\sc i}   &6158.187 & 10.741 & $-$0.441 &  7.61 & $-$3.96 &($-$7.23) & O~{\sc i} 6155--8 \\
 Fe~{\sc i}  &6159.368 &  4.607 & $-$1.970 &  8.28 & $-$4.61 & $-$7.77  & \\
 Na~{\sc i}  &6160.747 &  2.104 & $-$1.260 &  7.85 & $-$4.39 &($-$7.29) & Na~{\sc i} 6161\\
 Ca~{\sc i}  &6161.297 &  2.523 & $-$1.020 &  7.49 & $-$4.69 & $-$7.50  & \\
 Ca~{\sc i}  &6162.173 &  1.899 &  +0.100 &  7.82 & $-$5.07 & $-$7.59  & \\
\hline
\multicolumn{8}{c}{[7110--7121~$\rm\AA$ fitting]}\\
 Ni~{\sc i}  &7110.892 &  1.935 & $-$2.880 &  7.72 & $-$6.30 & $-$7.85  &(adjusted $gf$) \\
 C~{\sc i}   &7111.472 &  8.640 & $-$1.240 & (7.64)&($-$5.02)&($-$7.24) &C~{\sc i} 7111, (adjusted $gf$) \\
 C~{\sc i}   &7113.178 &  8.647 & $-$0.800 & (7.64)&($-$5.02)&($-$7.23) &C~{\sc i} 7113, (adjusted $gf$) \\
 C~{\sc i}   &7115.172 &  8.643 & $-$0.960 & (7.64)&($-$5.02)&($-$7.23) &C~{\sc i} 7115, (adjusted $gf$) \\
 C~{\sc i}   &7115.182 &  8.640 & $-$1.550 & (7.64)&($-$5.02)&($-$7.24) &C~{\sc i} 7115  \\
 C~{\sc i}   &7116.991 &  8.647 & $-$0.910 & (7.64)&($-$5.02)&($-$7.23) &C~{\sc i} 7116  \\
 Fe~{\sc i}  &7118.119 &  5.009 & $-$1.390 &  8.69 & $-$5.22 & $-$7.75  &(adjusted $gf$) \\
 C~{\sc i}   &7119.656 &  8.643 & $-$1.130 & (7.64)&($-$5.02)&($-$7.23) &C~{\sc i} 7119, (adjusted $gf$) \\
 Fe~{\sc i}  &7120.022 &  4.558 & $-$1.911 &  8.55 & $-$5.28 & $-$7.68  &(adjusted $gf$) \\
\hline
\multicolumn{8}{c}{[8677--8697~$\rm\AA$ fitting]}\\
 S~{\sc i}   &8678.927 &  7.867 & $-$1.000 &  7.61 & $-$4.41 &($-$7.30) & \\
 S~{\sc i}   &8679.620 &  7.867 & $-$0.410 &  7.61 & $-$4.41 &($-$7.30) & \\
 Si~{\sc i}  &8680.080 &  5.862 & $-$1.000 & (7.47)&($-$5.00)&($-$7.23) & \\
 N~{\sc i}   &8680.282 & 10.336 & +0.236 &  8.62 & $-$5.51 &($-$7.63) & N~{\sc i} 8680\\
 S~{\sc i}   &8680.411 &  7.867 & $-$0.210 &  7.61 & $-$4.41 &($-$7.30) & \\
 N~{\sc i}   &8683.403 & 10.330 & $-$0.045 &  8.62 & $-$5.51 &($-$7.64) & N~{\sc i} 8683\\
 N~{\sc i}   &8686.149 & 10.326 & $-$0.448 &  8.62 & $-$5.51 &($-$7.64) & N~{\sc i} 8686\\
 Si~{\sc i}  &8686.352 &  6.206 & $-$0.700 & (7.47)&($-$5.00)&($-$7.23) &(adjusted $gf$) \\
 Fe~{\sc i}  &8688.621 &  2.176 & $-$1.212 &  7.28 & $-$6.24 & $-$7.85  & \\
 Fe~{\sc i}  &8689.857 &  5.105 & $-$1.948 &  8.07 & $-$5.77 & $-$7.73  &(adjusted $gf$) \\
 Si~{\sc i}  &8690.061 &  5.613 & $-$1.980 & (7.47)&($-$5.28)&($-$7.32) & \\
 S~{\sc i}   &8693.137 &  7.870 & $-$1.370 &  7.62 & $-$4.41 &($-$7.30) & \\
 S~{\sc i}   &8693.931 &  7.870 & $-$0.510 &  7.62 & $-$4.41 &($-$7.30) & \\
 S~{\sc i}   &8694.626 &  7.870 &  +0.080 &  7.62 & $-$4.41 &($-$7.30) & \\
\hline
\end{tabular}
\end{center}
Note. 
All data are were taken from Kurucz \& Bell's (1995) compilation
as far as available, though empirically adjusted values (solar $gf$ values) 
were used for several lines remarked as ``adjusted'' in Column (8).
The meanings of Columns (1)--(4) are self-explanatory.
In Columns (5)--(7) are given the damping parameters in the c.g.s. unit:\\
Gammar is the radiation damping constant (s$^{-1}$), $\log\gamma_{\rm rad}$.
Gammas is the Stark damping width per electron density (cm$^{-3}$)
at $10^{4}$ K, $\log(\gamma_{\rm e}/N_{\rm e})$.
Gammaw is the van der Waals damping width per hydrogen density (cm$^{-3}$)
at $10^{4}$ K, $\log(\gamma_{\rm w}/N_{\rm H})$. 
Note that the values in parentheses are the default damping parameters 
computed within the Kurucz's (1993) WIDTH9 program (cf. Leusin \& Topil'skaya 1987),
because the damping data for these lines were unavailable in 
Kurucz \& Bell's (1995) database. 
\end{minipage}
\end{table*}

\setcounter{figure}{0}
\begin{figure*}
\begin{minipage}{120mm}
\includegraphics[width=12.0cm]{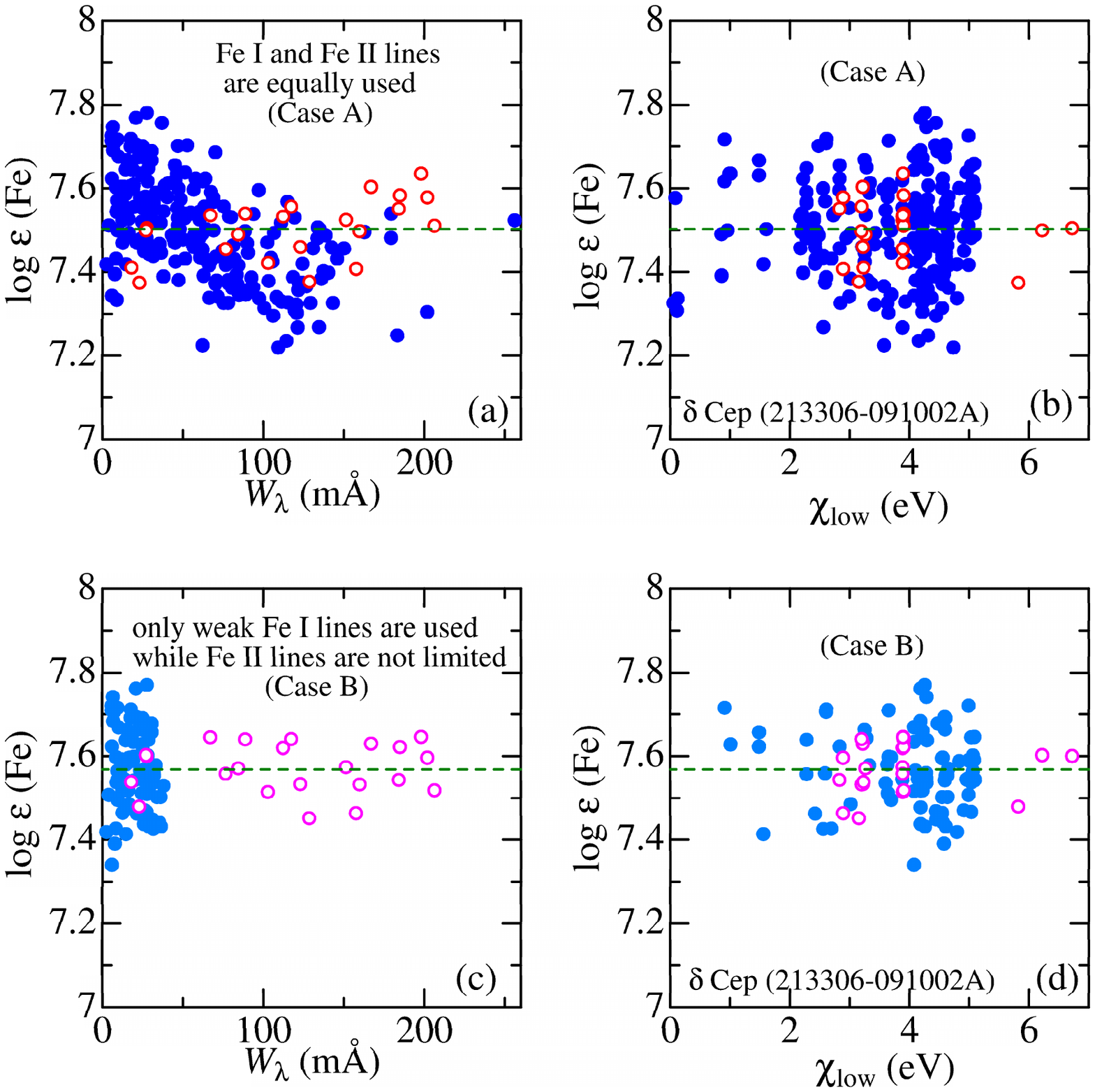}
\caption{Line-by-line Fe abundances (corresponding to the finally 
established atmospheric parameters), plotted against equivalent
widths (left panels) and lower excitation potentials (right panels),
for the case of $\delta$~Cep (213306-091002A; the first 
spectrum for this star).
The upper two panels (a, b) are for Case (A) where lines of $w < 200$~m$\rm\AA$ 
($w$ is the reduced equivalent width: $w \equiv W_{\lambda}\cdot (5000/\lambda)$) 
were equally used for both Fe~{\sc i} and Fe~{\sc ii} lines, while the lower 
two panels (c, d) correspond to Case (B) where lines were screened with the 
criterion of $w < 30$~m$\rm\AA$ (Fe~{\sc i}) and $w < 200$~m$\rm\AA$ (Fe~{\sc ii}) 
(i.e., only weak Fe~{\sc i} lines were used). 
The results for Fe~{\sc i} and Fe~{\sc ii} lines are distinguished by filled 
and open symbols, respectively, and the average Fe abundance is indicated
by the horizontal dashed line.
}
\label{fig1}
\end{minipage}
\end{figure*}

\setcounter{figure}{1}
\begin{figure*}
\begin{minipage}{150mm}
\includegraphics[width=15.0cm]{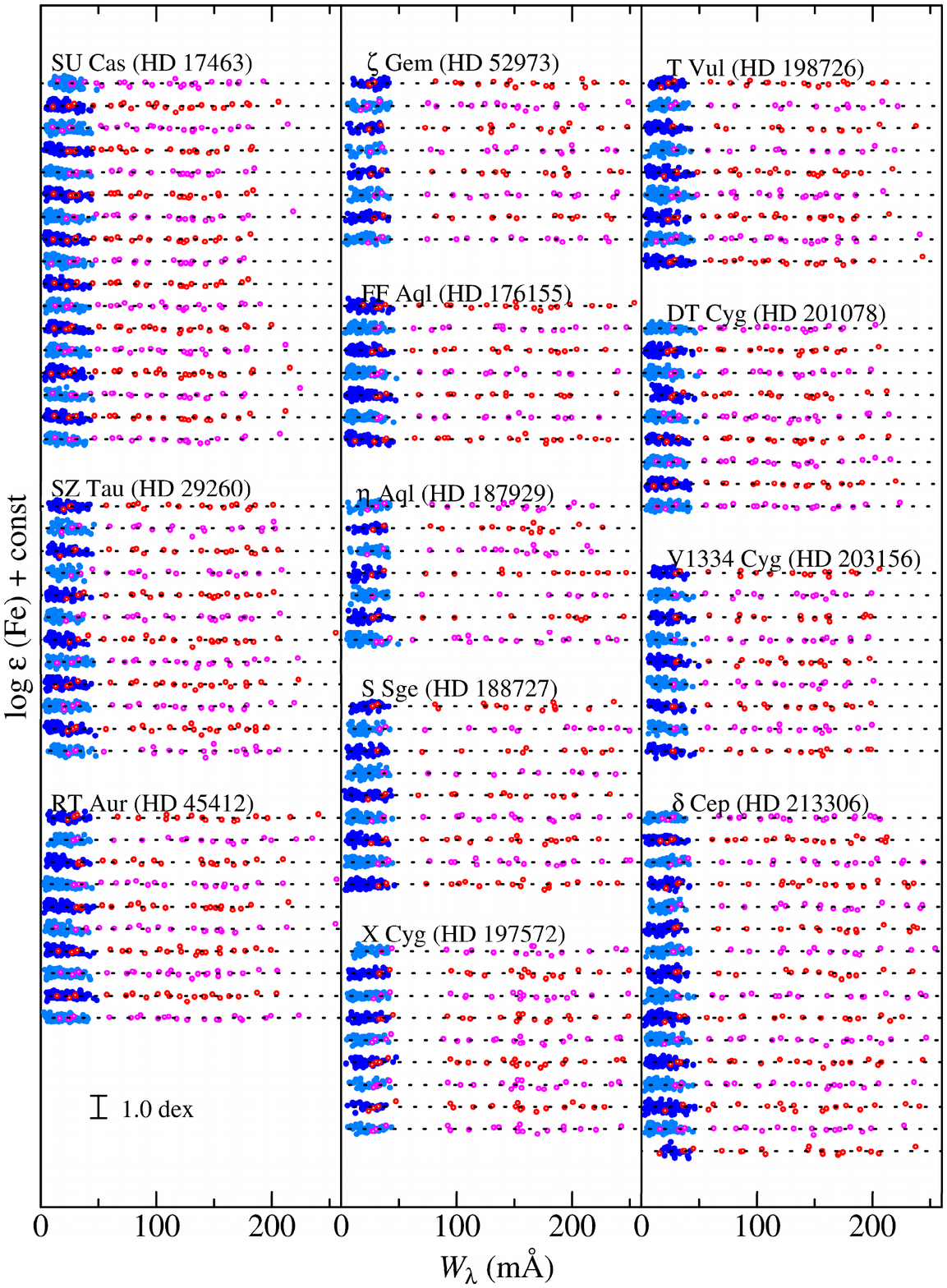}
\caption{Fe abundance vs. equivalent width relations 
corresponding to the finally established atmospheric parameters of 
$T_{\rm eff}$, $\log g$, and $v_{\rm t}$ for each of the 122 spectra,
being arranged according to the time sequence in the downward
direction for each star (just as in Table 2).
The filled and open symbols correspond to Fe~{\sc i} and Fe~{\sc ii} 
lines, respectively. The results for each stars are shown relative to 
the mean abundances indicated by the horizontal dotted lines, and 
vertically shifted by 1.0 relative to the adjacent ones.}
\label{fig2}
\end{minipage}
\end{figure*}

\setcounter{figure}{2}
\begin{figure*}
\begin{minipage}{150mm}
\includegraphics[width=15.0cm]{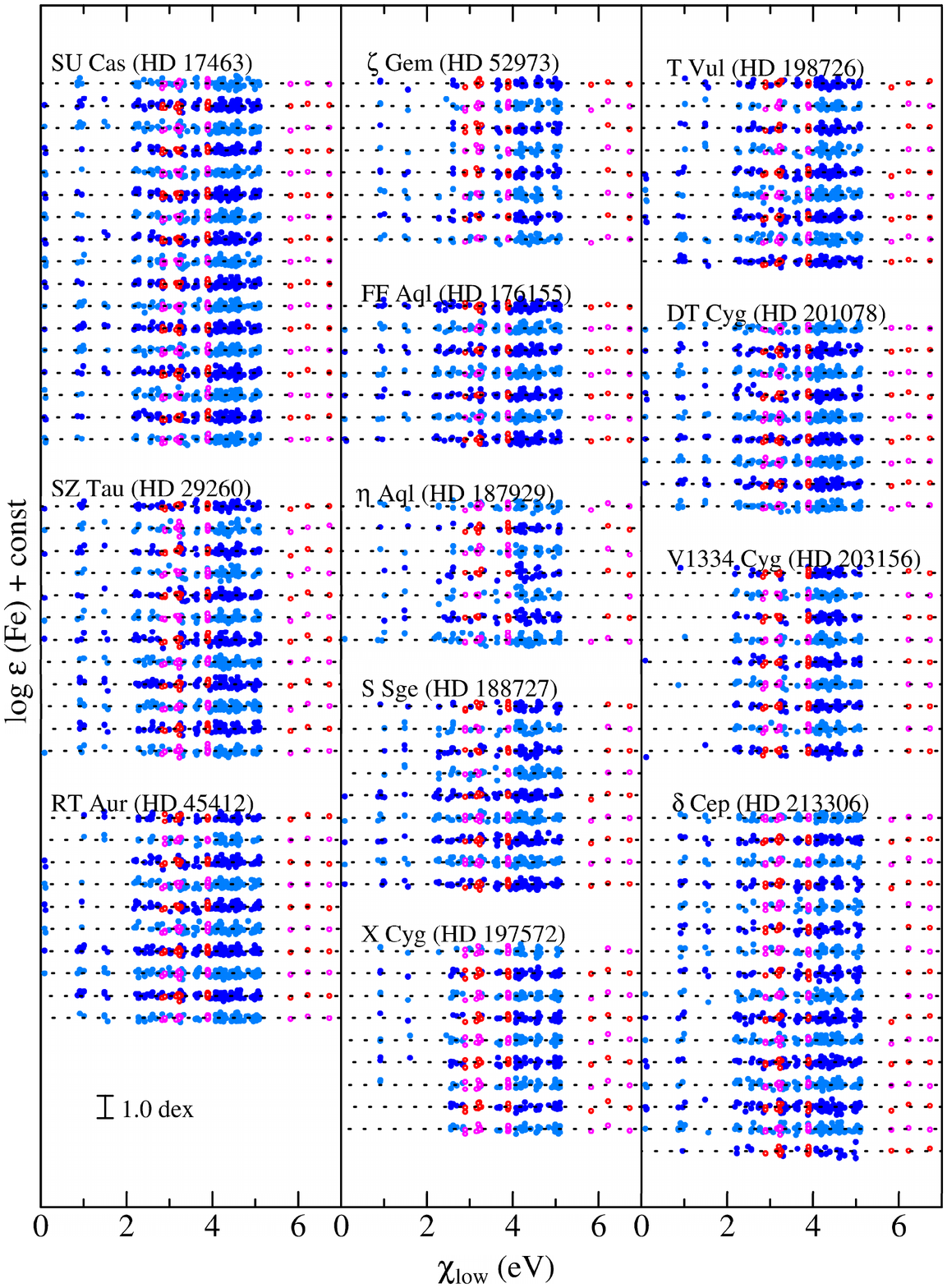}
\caption{Fe abundance vs. lower excitation potential relation
corresponding to the finally established atmospheric parameters
for each of the 122 spectra. Otherwise, the same as in Fig. 2.}
\label{fig3}
\end{minipage}
\end{figure*}

\setcounter{figure}{3}
\begin{figure*}
\begin{minipage}{150mm}
\includegraphics[width=15.0cm]{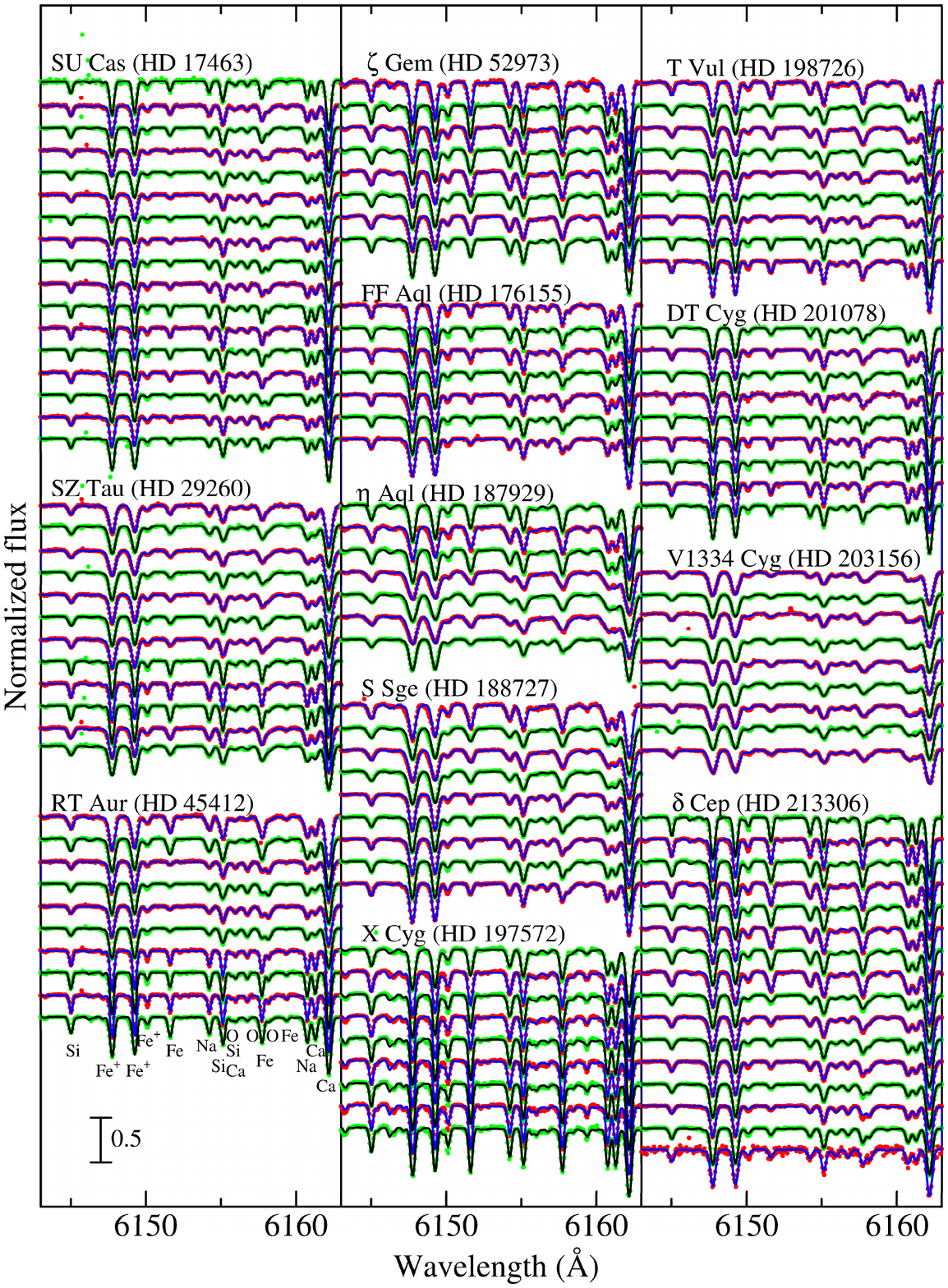}
\caption{Synthetic spectrum fitting in the 6143--6163~$\rm\AA$
region accomplished by varying the abundances of 
O, Na, Si, Ca, and Fe, along with the macrobroadening 
velocity ($v_{\rm M}$) and the wavelength shift (radial velocity).
The best-fit theoretical spectra are shown by solid lines, 
while the observed data are plotted by symbols, where
the wavelength scale of the stellar spectrum has been adjusted
to the laboratory frame.  
In each panel, the spectra are arranged according to the 
time sequence in the downward direction for each star 
(just as in Table 2), and each spectrum is vertically shifted by 
0.25 relative to the adjacent one. The important lines (cf. Table 3) 
are identified in the lowest spectrum of the left panel.
}
\label{fig4}
\end{minipage}
\end{figure*}

\setcounter{figure}{4}
\begin{figure*}
\begin{minipage}{150mm}
\includegraphics[width=15.0cm]{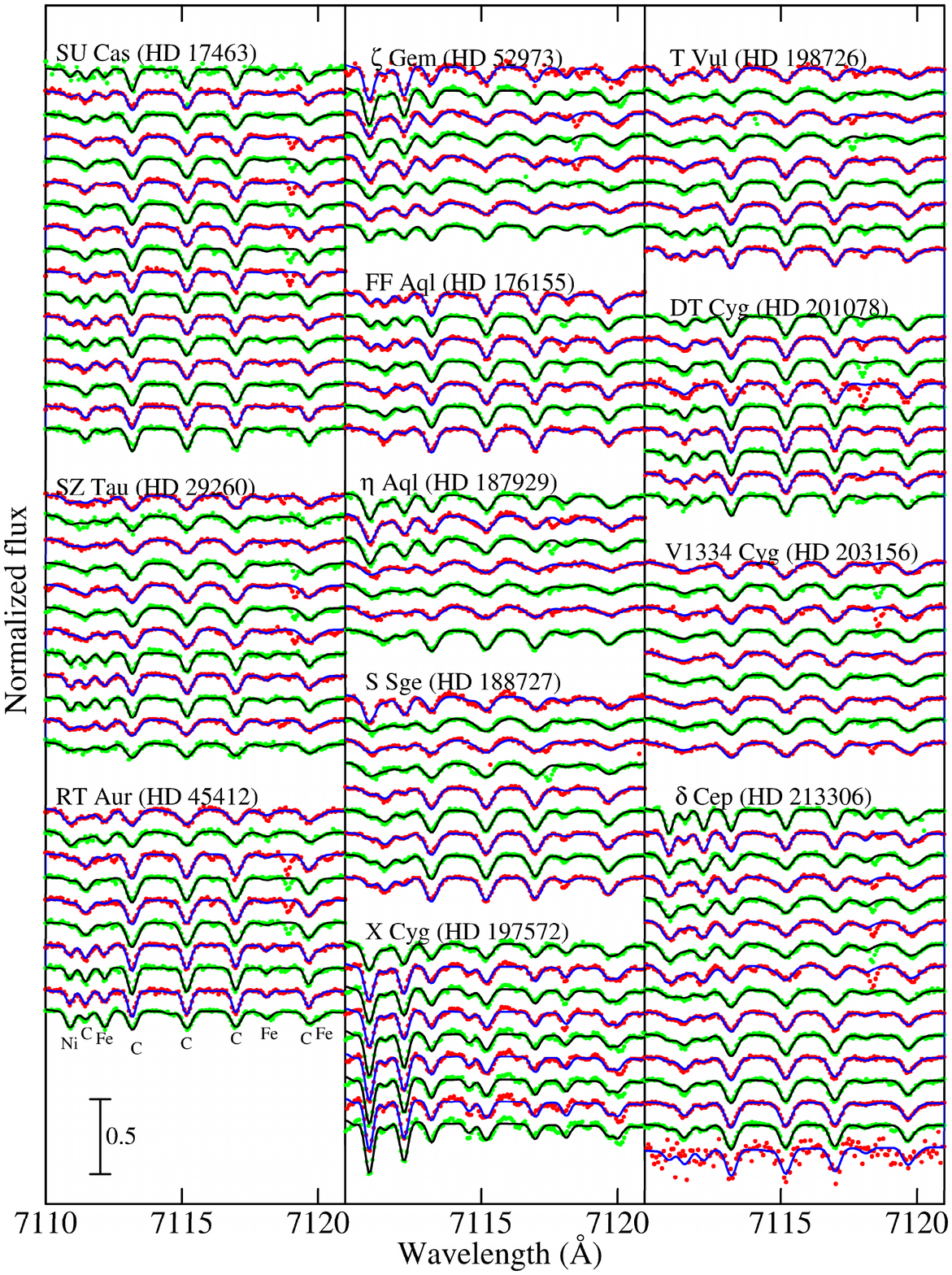}
\caption{
Synthetic spectrum fitting in the 7110--7121~$\rm\AA$
region accomplished by varying the abundances of 
C, Fe, and Ni. Each spectrum is vertically shifted by 0.15 
relative to the adjacent one. Otherwise, the same as in Fig. 4.}
\label{fig5}
\end{minipage}
\end{figure*}

\setcounter{figure}{5}
\begin{figure*}
\begin{minipage}{150mm}
\includegraphics[width=15.0cm]{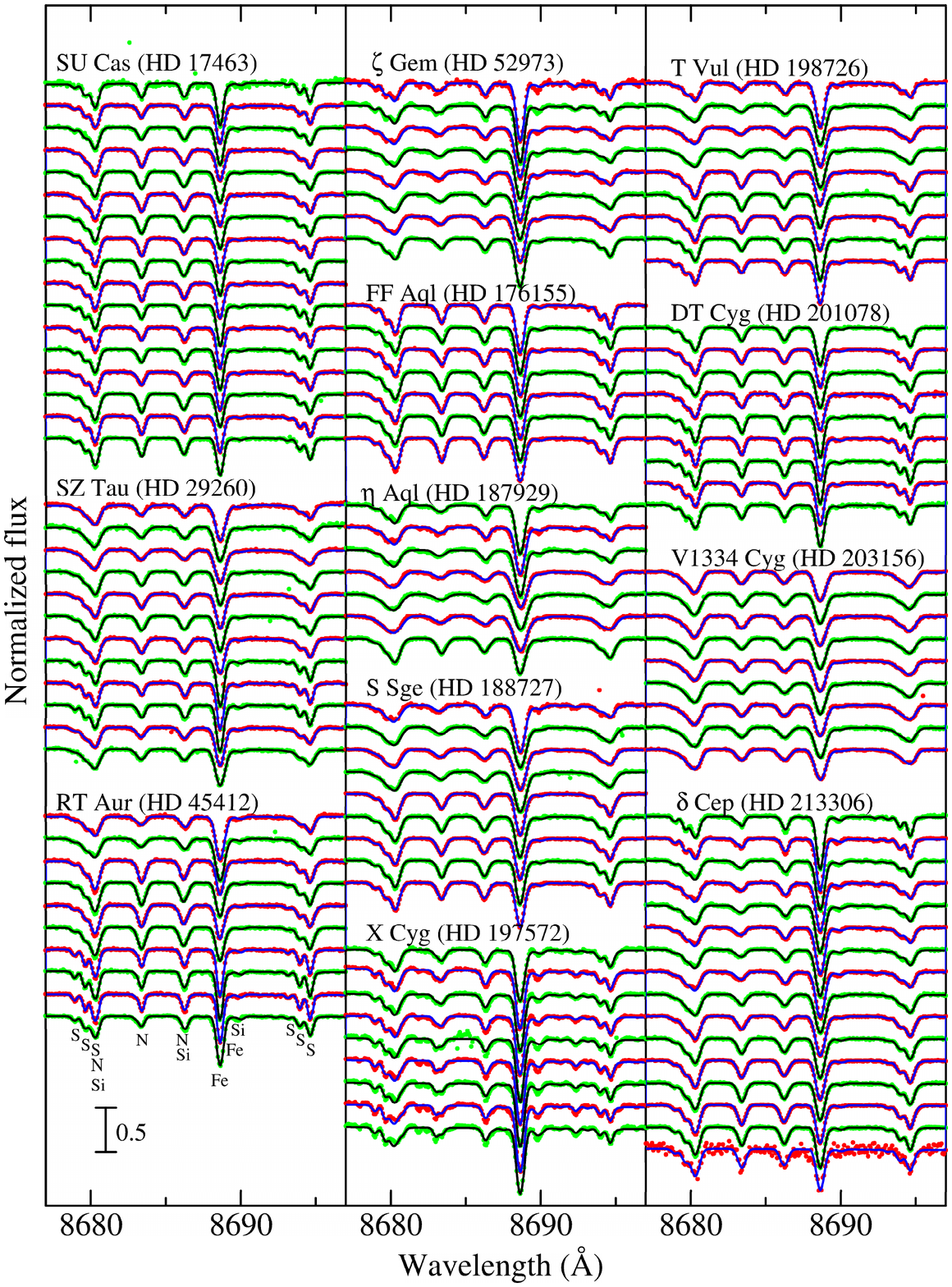}
\caption{Synthetic spectrum fitting in the 8677--8697~$\rm\AA$
region accomplished by varying the abundances of 
N, Si, S, and Fe. Each spectrum is vertically shifted by 0.25 
relative to the adjacent one. Otherwise, the same as in Fig. 4.}
\label{fig6}
\end{minipage}
\end{figure*}

\setcounter{figure}{6}
\begin{figure*}
\begin{minipage}{100mm}
\begin{center}
\includegraphics[width=10.0cm]{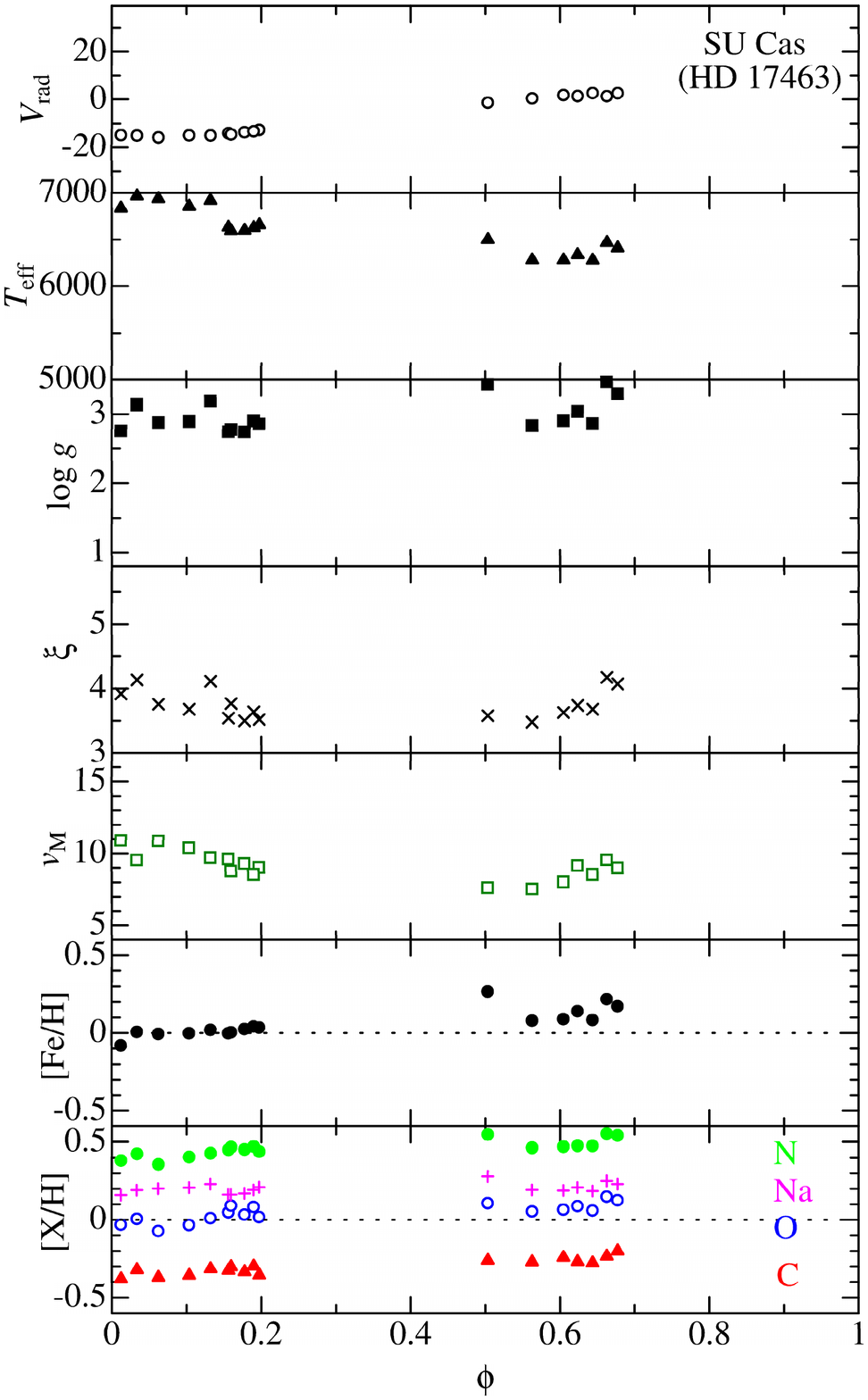}
\caption{Results of the radial velocity, atmospheric parameters, 
and elemental abundances derived from each of the 17 spectra 
of SU~Cas, plotted against the pulsation phase. Shown in these
7 panels are (from top to bottom): (a) heliocentric
radial velocity (km~s$^{-1}$), (b) effective temperature (K),
(c) logarithmic surface gravity (cm~s$^{-2}$), (d) microturbulence
(km~s$^{-1}$), (e) macrobroadening velocity (km~s$^{-1}$), 
(f) [Fe/H] (dex), and (g) [X/H] (dex) where X is C (filled triangles),
N (filled circles), O (open circles), and Na (greek crosses).
 }
\label{fig7}
\end{center}
\end{minipage}
\end{figure*}

\setcounter{figure}{7}
\begin{figure*}
\begin{minipage}{100mm}
\begin{center}
\includegraphics[width=10.0cm]{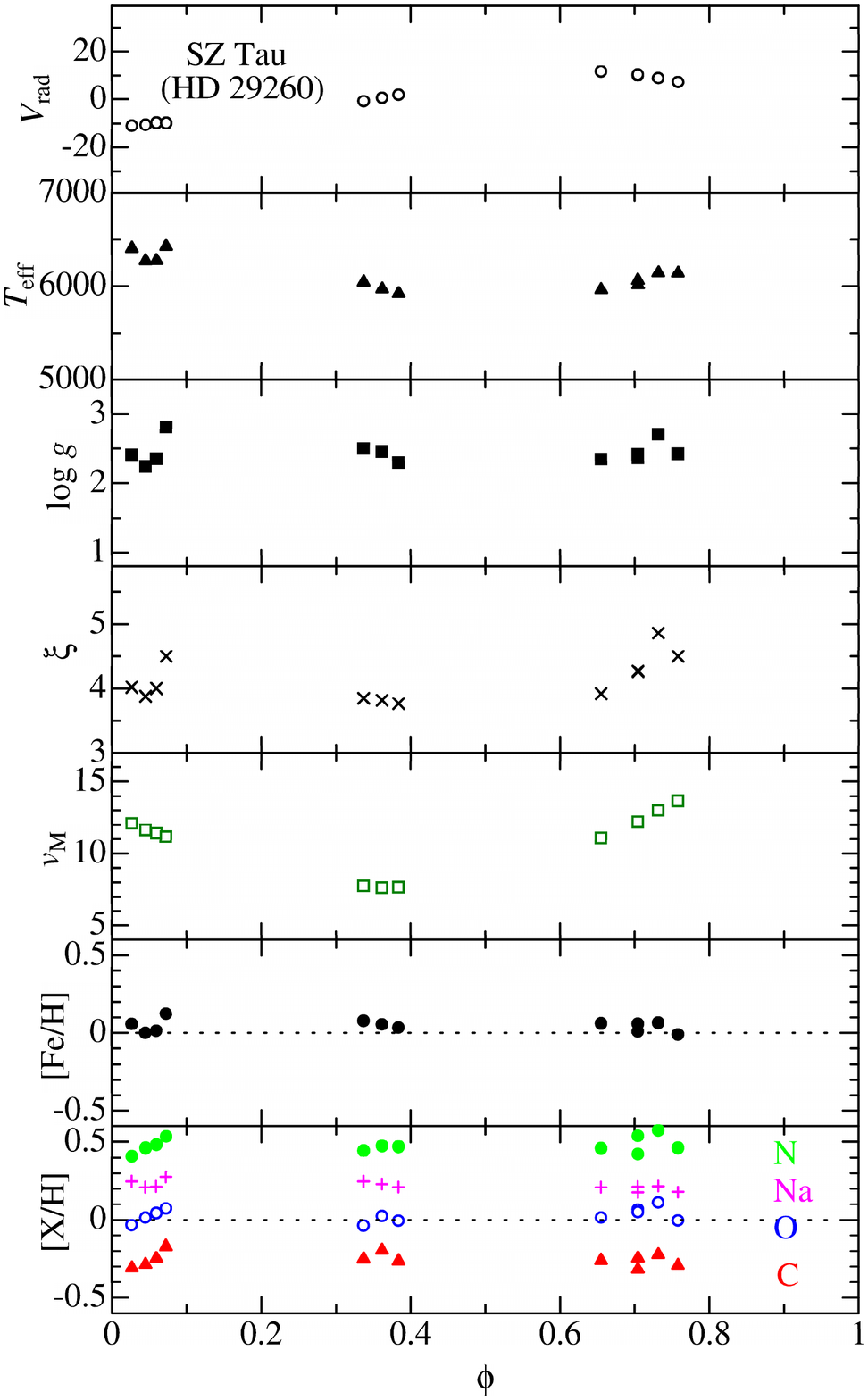}
\caption{Results of the radial velocity, atmospheric parameters, 
and elemental abundances derived from each of the 12 spectra 
of SZ~Tau, plotted against the pulsation phase. Otherwise, 
the same as in Fig. 7.}
\label{fig8}
\end{center}
\end{minipage}
\end{figure*}

\setcounter{figure}{8}
\begin{figure*}
\begin{minipage}{100mm}
\begin{center}
\includegraphics[width=10.0cm]{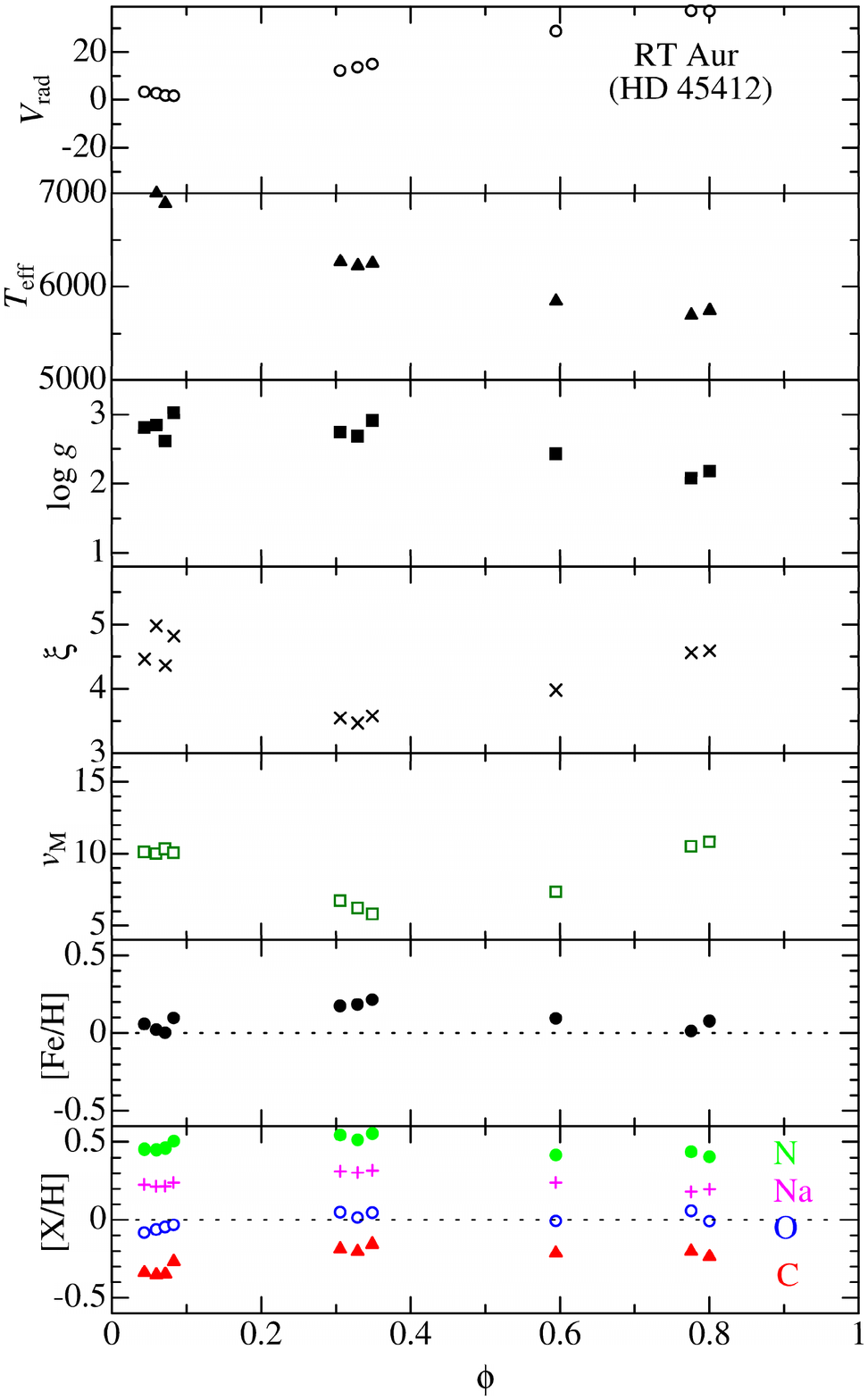}
\caption{Results of the radial velocity, atmospheric parameters, 
and elemental abundances derived from each of the 10 spectra 
of RT Aur, plotted against the pulsation phase. Otherwise, 
the same as in Fig. 7.}
\label{fig9}
\end{center}
\end{minipage}
\end{figure*}

\setcounter{figure}{9}
\begin{figure*}
\begin{minipage}{100mm}
\begin{center}
\includegraphics[width=10.0cm]{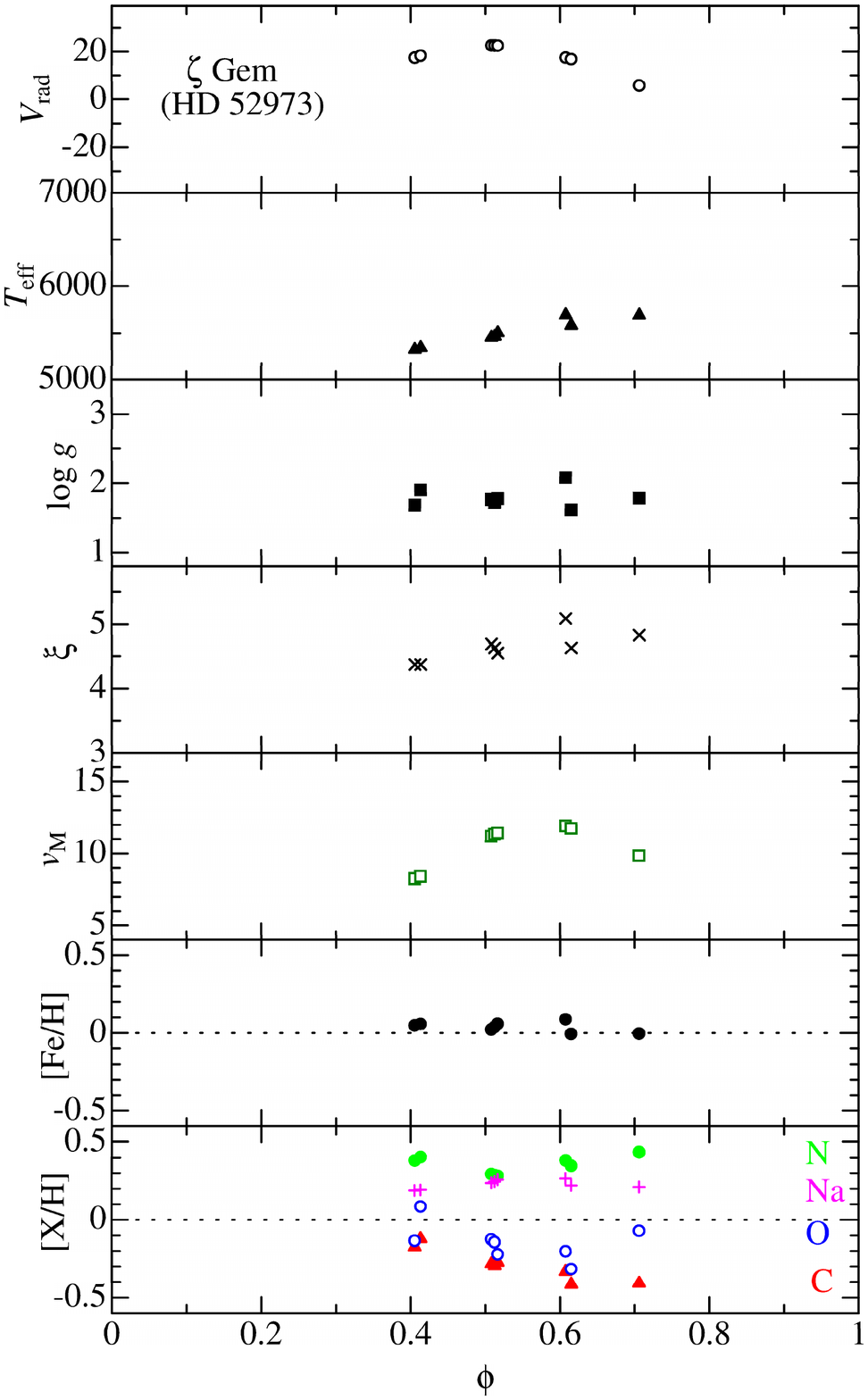}
\caption{Results of the radial velocity, atmospheric parameters, 
and elemental abundances derived from each of the 8 spectra 
of $\zeta$~Gem, plotted against the pulsation phase. Otherwise, 
the same as in Fig. 7.}
\label{fig10}
\end{center}
\end{minipage}
\end{figure*}

\setcounter{figure}{10}
\begin{figure*}
\begin{minipage}{100mm}
\begin{center}
\includegraphics[width=10.0cm]{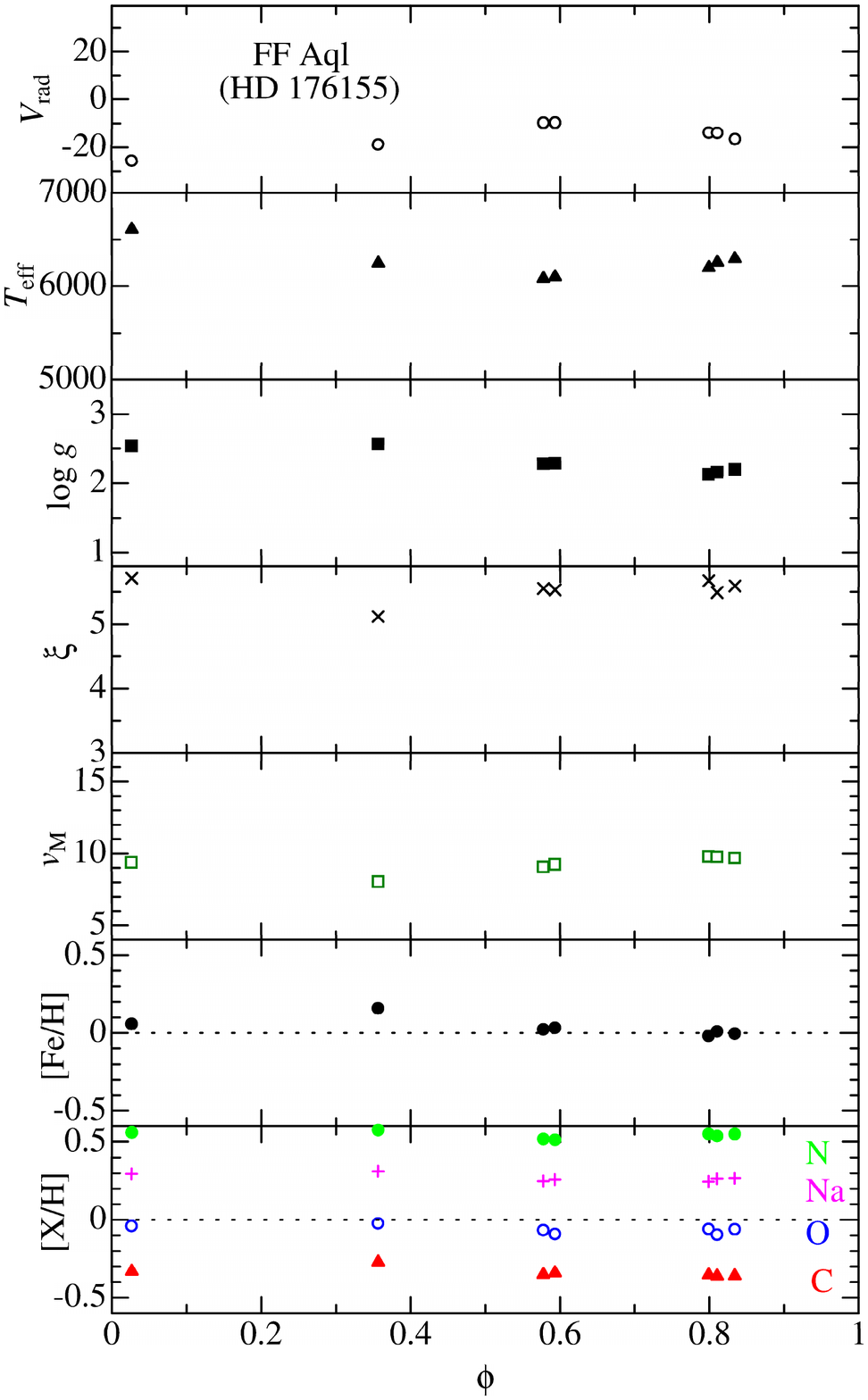}
\caption{Results of the radial velocity, atmospheric parameters, 
and elemental abundances derived from each of the 7 spectra 
of FF~Aql, plotted against the pulsation phase. Otherwise, 
the same as in Fig. 7.}
\label{fig11}
\end{center}
\end{minipage}
\end{figure*}

\setcounter{figure}{11}
\begin{figure*}
\begin{minipage}{100mm}
\begin{center}
\includegraphics[width=10.0cm]{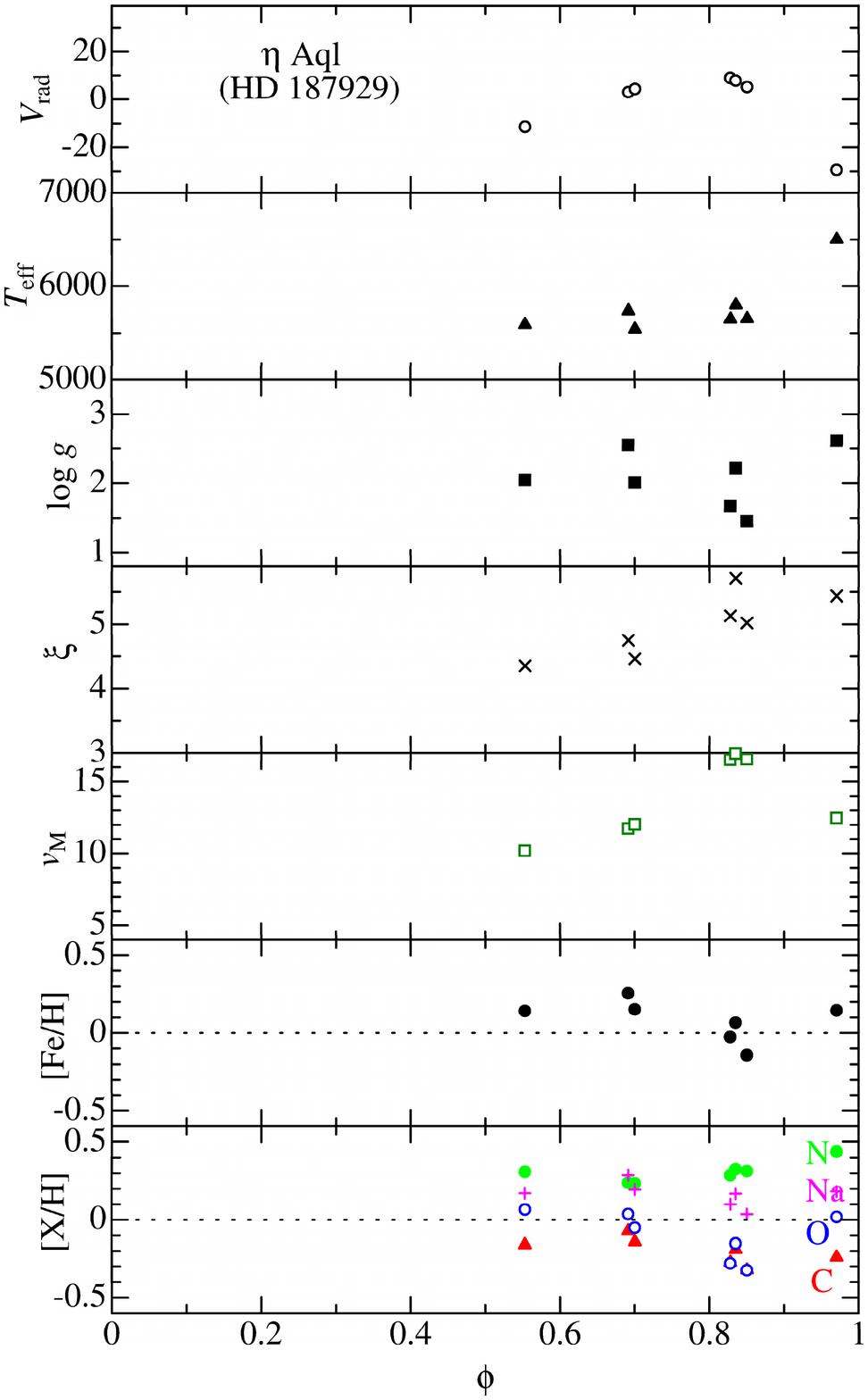}
\caption{Results of the radial velocity, atmospheric parameters, 
and elemental abundances derived from each of the 7 spectra 
of $\eta$~Aql, plotted against the pulsation phase. Otherwise, 
the same as in Fig. 7.}
\label{fig12}
\end{center}
\end{minipage}
\end{figure*}

\setcounter{figure}{12}
\begin{figure*}
\begin{minipage}{100mm}
\begin{center}
\includegraphics[width=10.0cm]{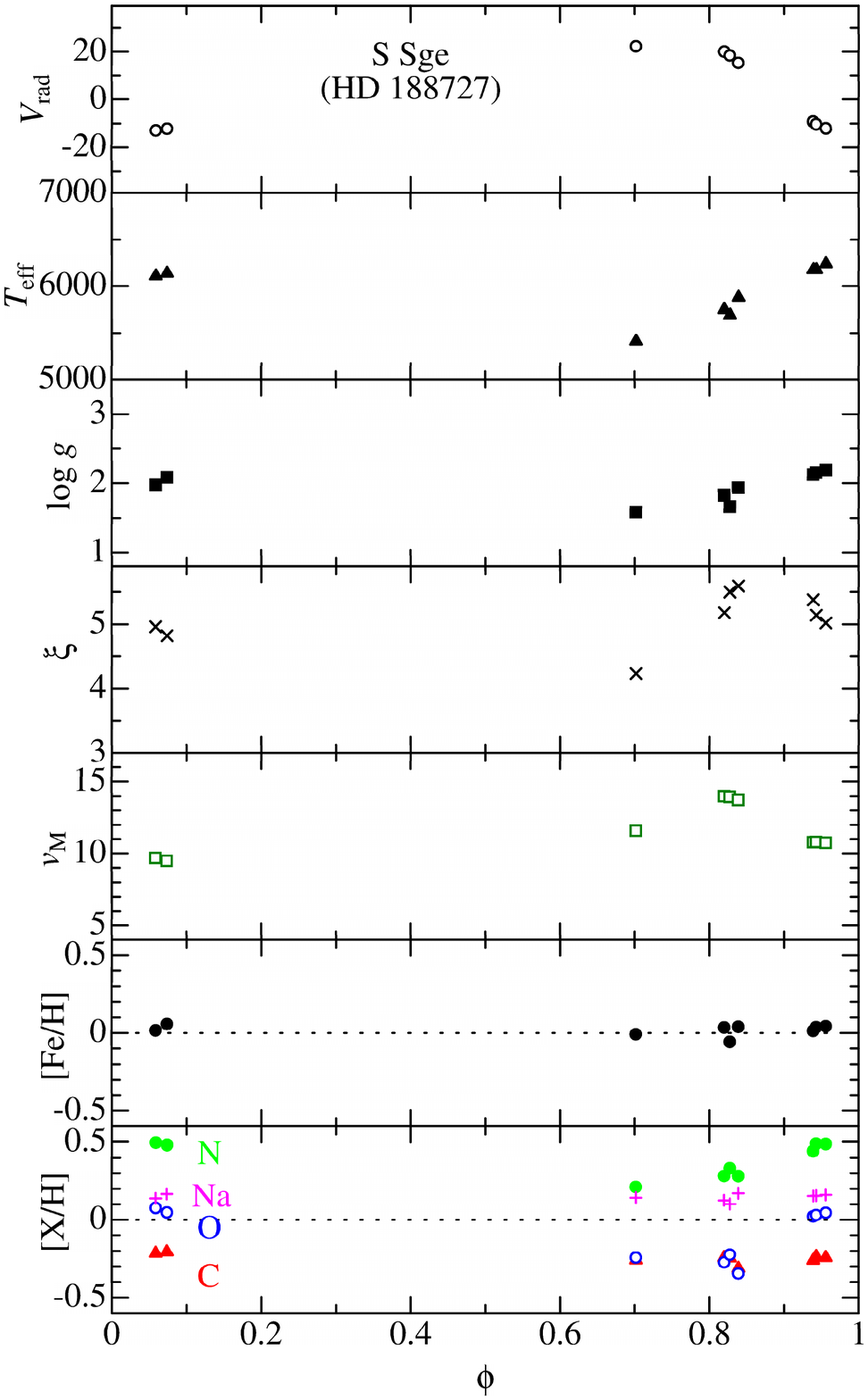}
\caption{Results of the radial velocity, atmospheric parameters, 
and elemental abundances derived from each of the 9 spectra 
of S~Sge, plotted against the pulsation phase. Otherwise, 
the same as in Fig. 7.}
\label{fig13}
\end{center}
\end{minipage}
\end{figure*}

\clearpage

\setcounter{figure}{13}
\begin{figure*}
\begin{minipage}{100mm}
\begin{center}
\includegraphics[width=10.0cm]{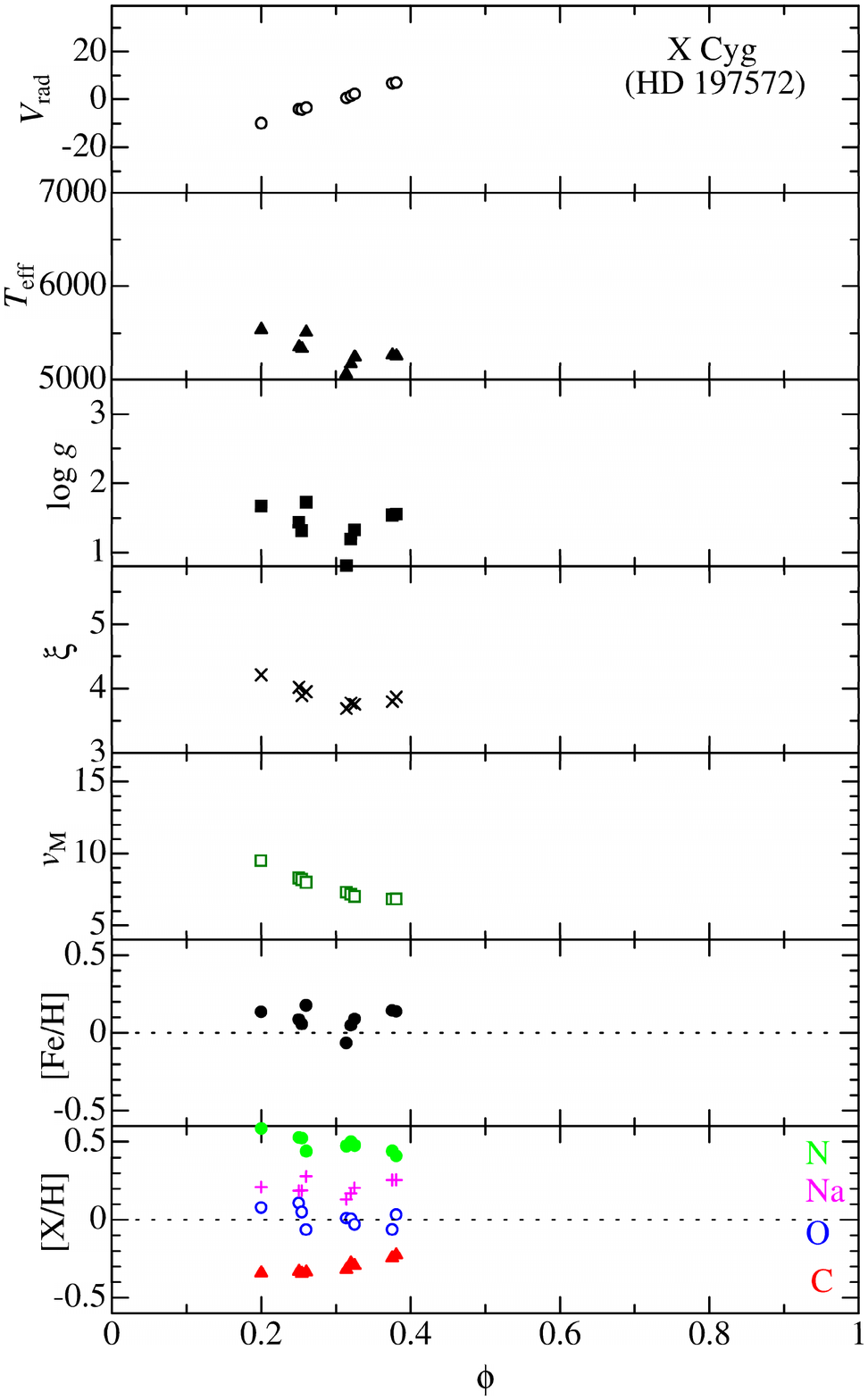}
\caption{Results of the radial velocity, atmospheric parameters, 
and elemental abundances derived from each of the 9 spectra 
of X~Cyg, plotted against the pulsation phase. Otherwise, 
the same as in Fig. 7.}
\label{fig14}
\end{center}
\end{minipage}
\end{figure*}

\setcounter{figure}{14}
\begin{figure*}
\begin{minipage}{100mm}
\begin{center}
\includegraphics[width=10.0cm]{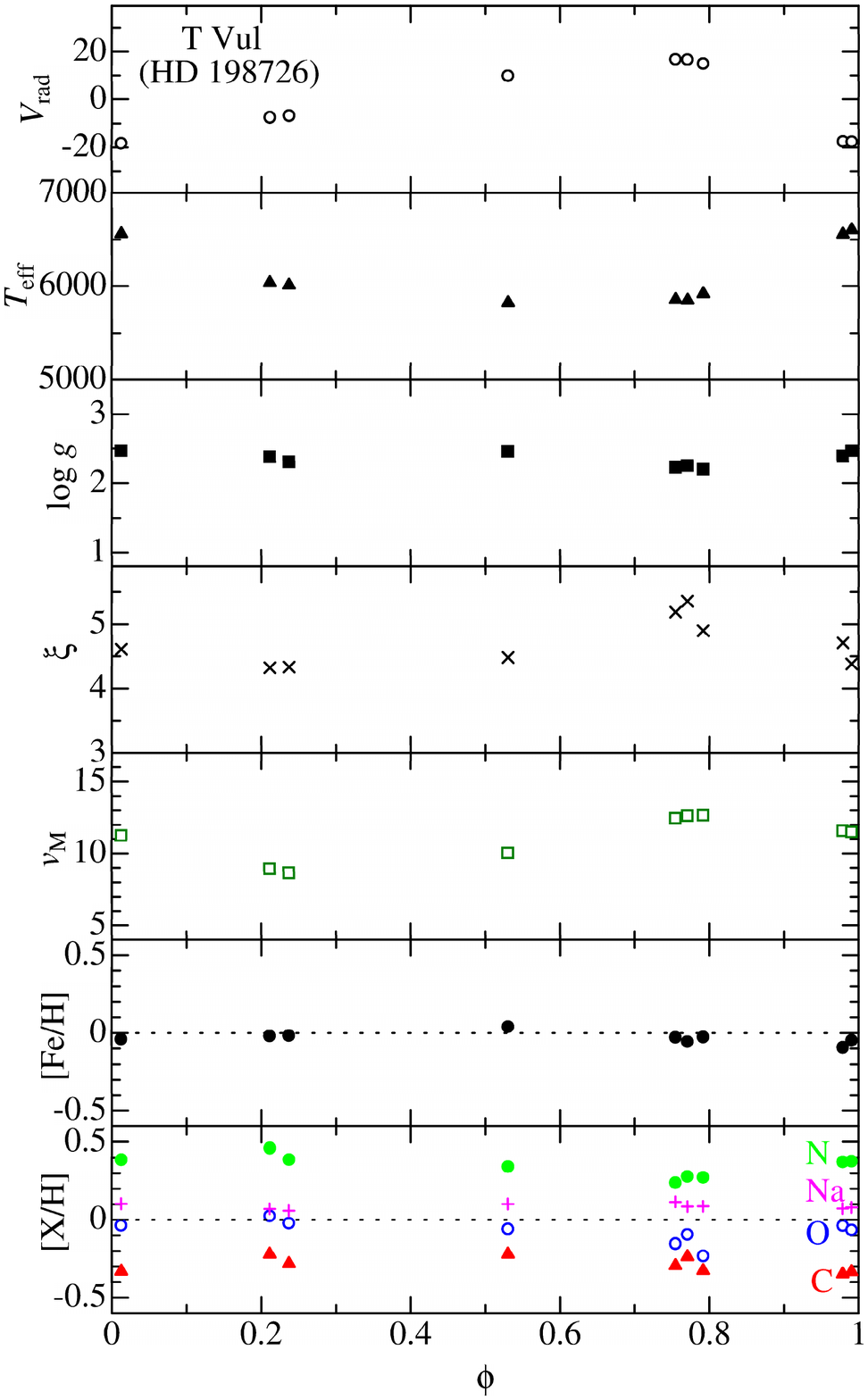}
\caption{Results of the radial velocity, atmospheric parameters, 
and elemental abundances derived from each of the 9 spectra 
of T~Vul, plotted against the pulsation phase. Otherwise, 
the same as in Fig. 7.}
\label{fig15}
\end{center}
\end{minipage}
\end{figure*}

\setcounter{figure}{15}
\begin{figure*}
\begin{minipage}{100mm}
\begin{center}
\includegraphics[width=10.0cm]{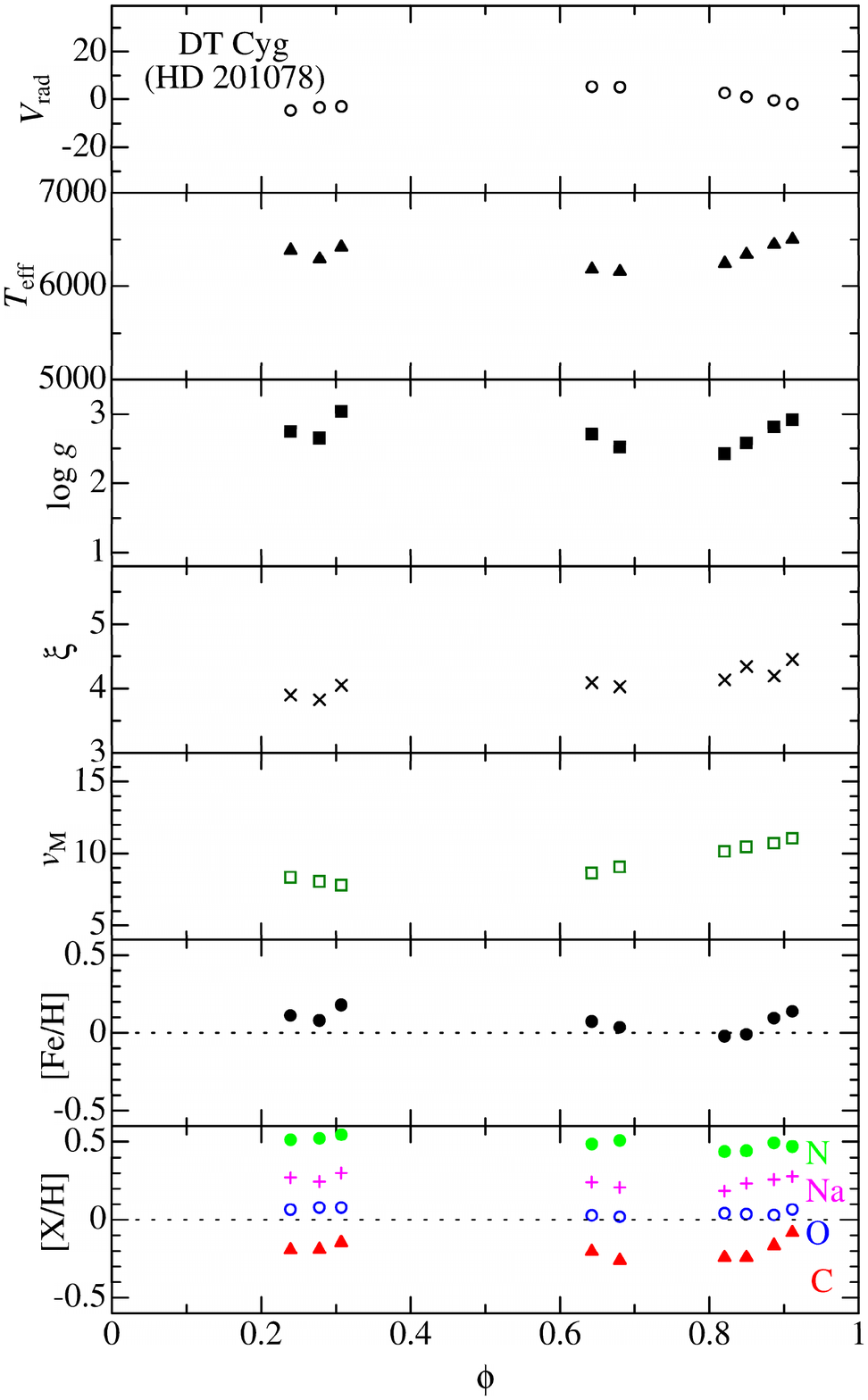}
\caption{Results of the radial velocity, atmospheric parameters, 
and elemental abundances derived from each of the 9 spectra 
of DT~Cyg, plotted against the pulsation phase. Otherwise, 
the same as in Fig. 7.}
\label{fig16}
\end{center}
\end{minipage}
\end{figure*}

\setcounter{figure}{16}
\begin{figure*}
\begin{minipage}{100mm}
\begin{center}
\includegraphics[width=10.0cm]{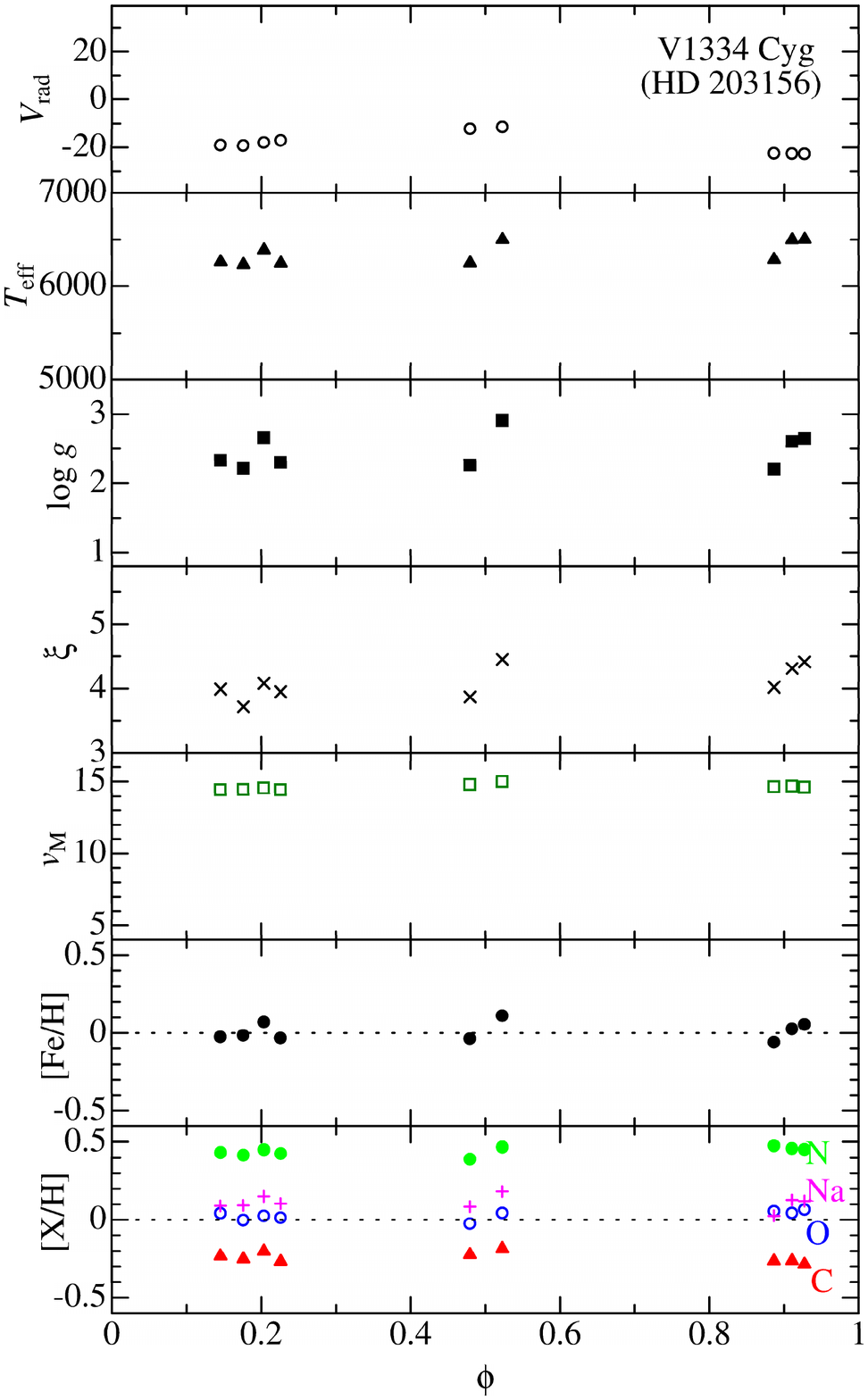}
\caption{Results of the radial velocity, atmospheric parameters, 
and elemental abundances derived from each of the 9 spectra 
of V1334~Cyg, plotted against the pulsation phase. Otherwise, 
the same as in Fig. 7.}
\label{fig17}
\end{center}
\end{minipage}
\end{figure*}

\setcounter{figure}{17}
\begin{figure*}
\begin{minipage}{100mm}
\begin{center}
\includegraphics[width=10.0cm]{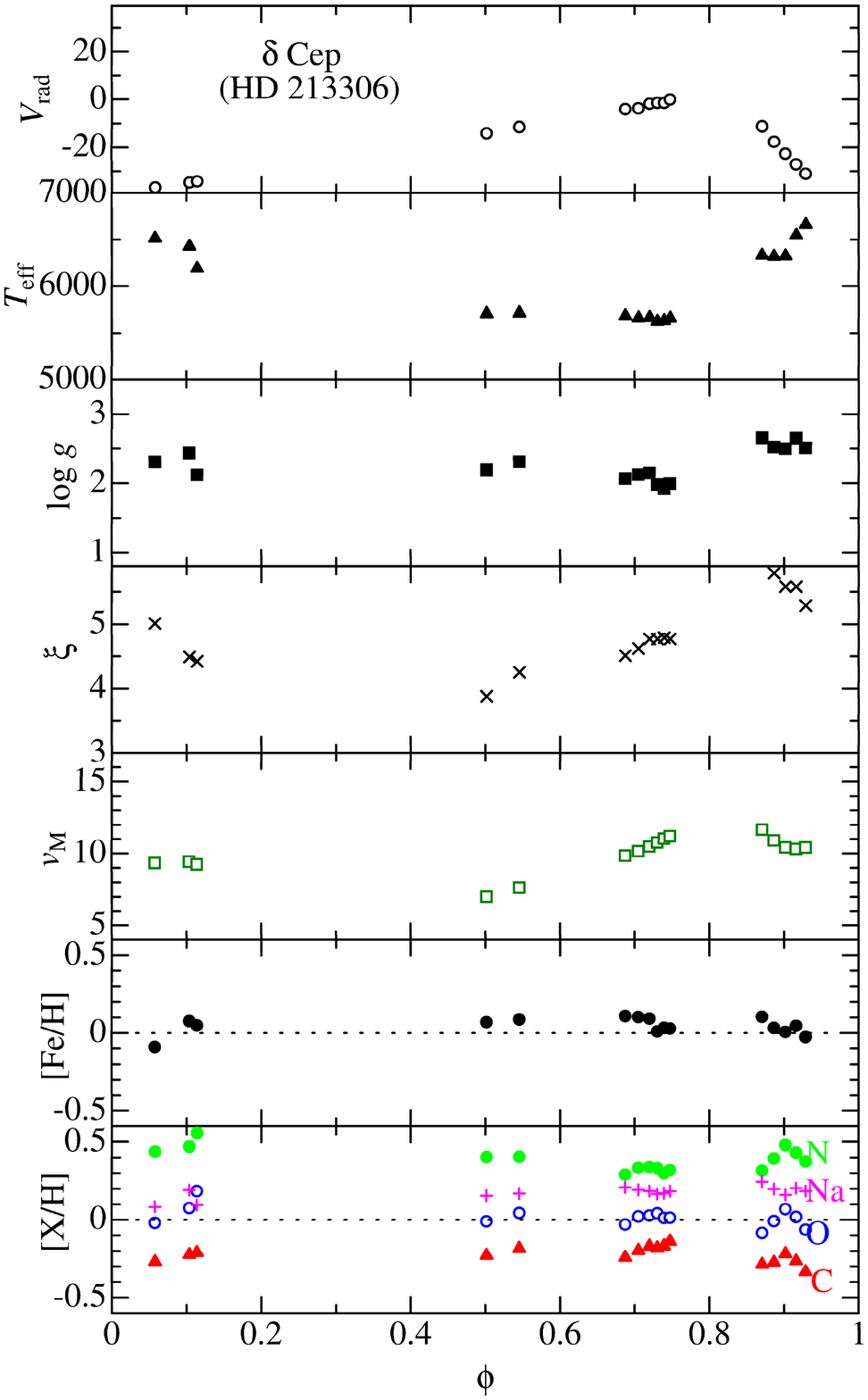}
\caption{Results of the radial velocity, atmospheric parameters, 
and elemental abundances derived from each of the 16 spectra 
of $\delta$~Cep, plotted against the pulsation phase. Otherwise, 
the same as in Fig. 7.}
\label{fig18}
\end{center}
\end{minipage}
\end{figure*}

\setcounter{figure}{18}
\begin{figure*}
\begin{minipage}{120mm}
\includegraphics[width=12.0cm]{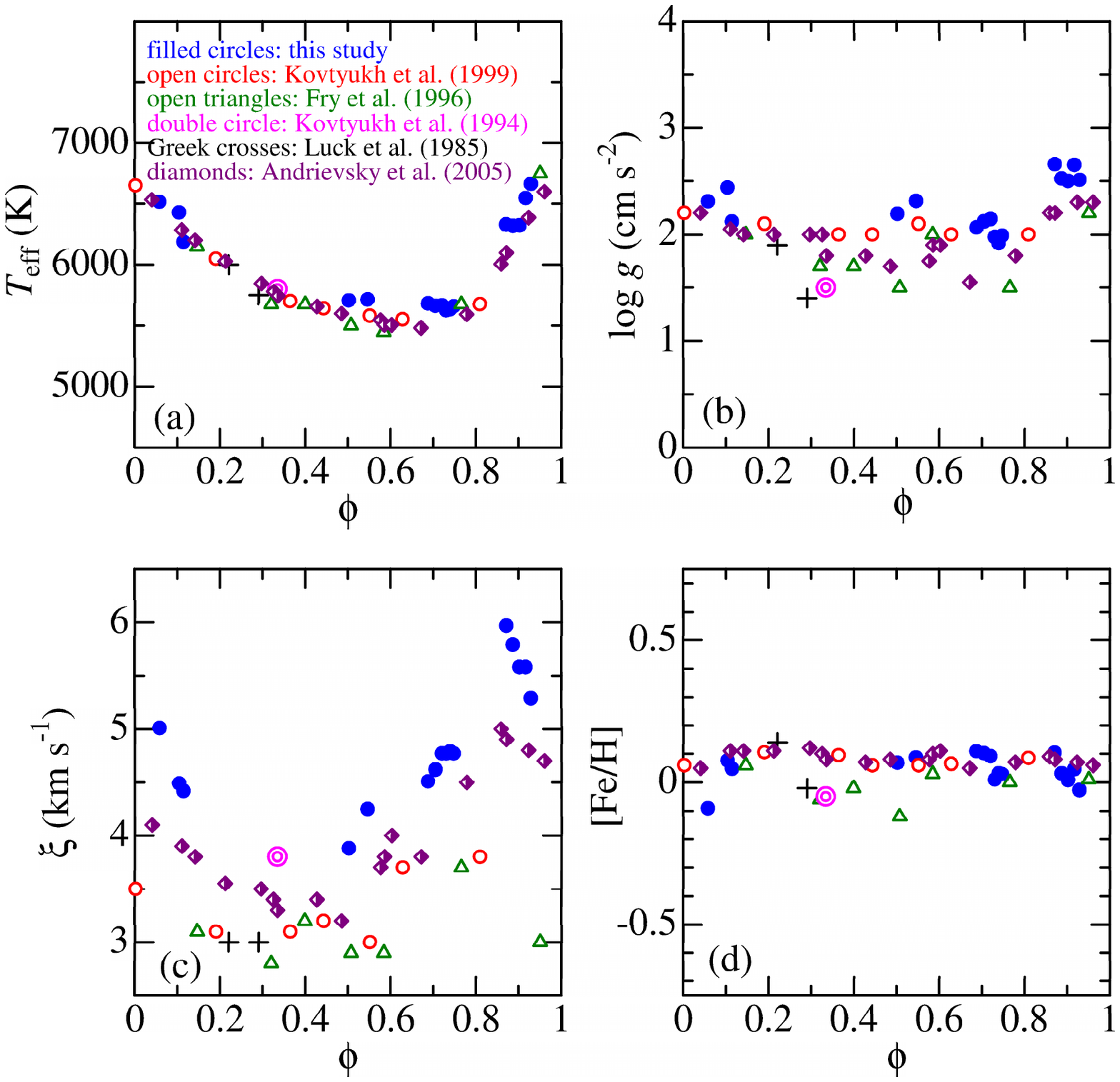}
\caption{Phase($\phi$)-dependence of the atmospheric parameters of 
$\delta$ Cep (HD~213306), where our results and the previous results of 
other similar ``spectroscopic'' determinations are overplotted with
different symbols: Filled circles --- this study, open circles ---
Kovtyukh \& Andrievsky (1999) (spectroscopic $T_{\rm eff}$ and non-standard 
results for $\log g$, $\xi$,and [Fe/H]), open triangles --- Fry \& Carney
(1997), double circle --- Kovtyukh et al. (1994) (note that their 
$T_{\rm eff}$ was determined from H$\alpha$ and colors), Greek crosses ---
Luck \& Lambert (1985) (spectroscopic $\log g$), diamonds ---
Andrievsky et al. (2005).
(a) $T_{\rm eff}$ vs. $\phi$, (b) $\log g$ vs. $\phi$, 
(c) $\xi$ vs. $\phi$, and (d) [Fe/H] vs. $\phi$. 
 }
\label{fig19}
\end{minipage}
\end{figure*}

\setcounter{figure}{19}
\begin{figure*}
\begin{minipage}{120mm}
\includegraphics[width=12.0cm]{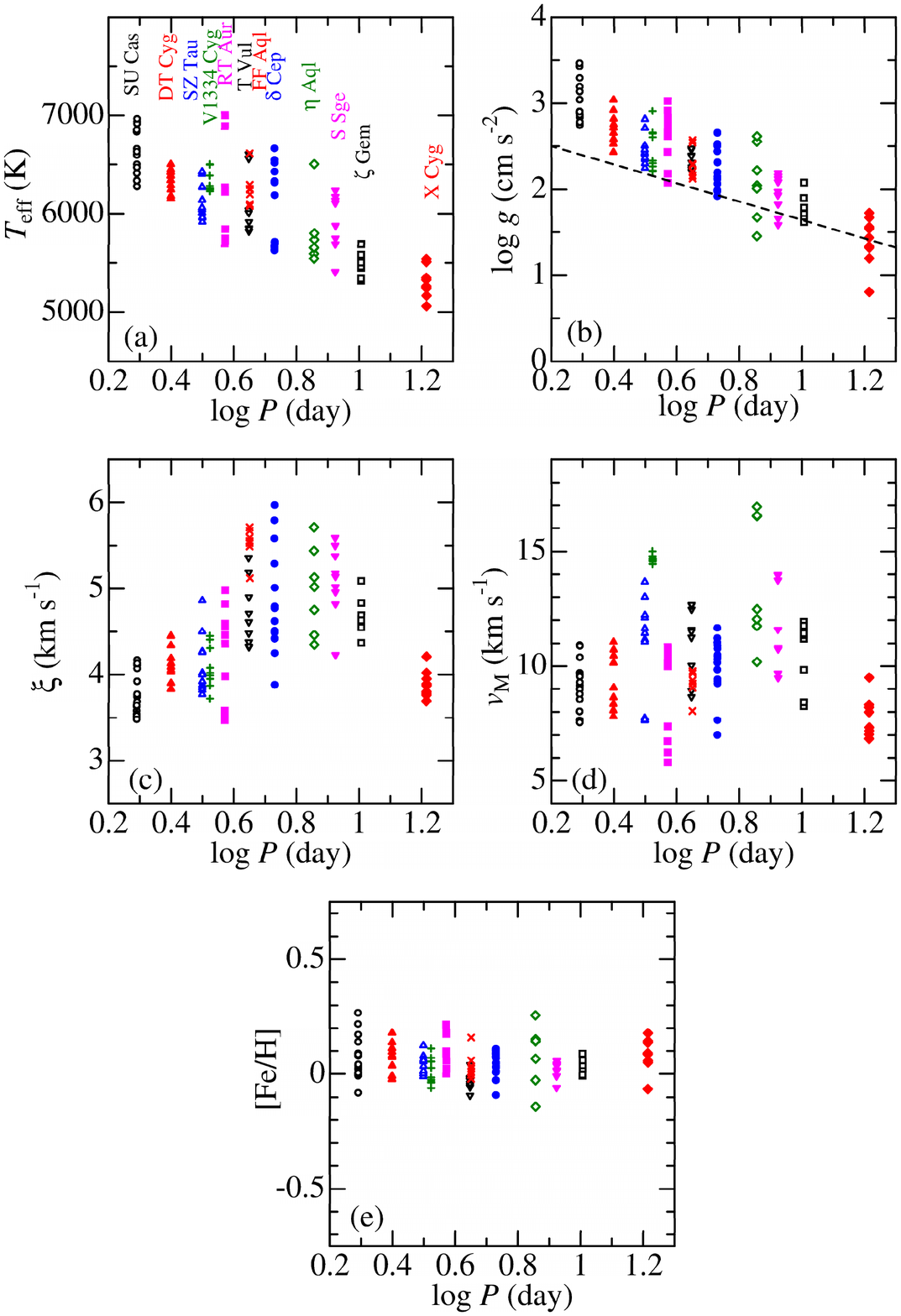}
\caption{Atmospheric parameters derived from each of the 122 spectra, 
plotted against the pulsation period ($\log P$): (a) effective 
temperature, (b) logarithmic surface gravity, (c) microturbulence, 
(d) macrobroadening velocity, and (e) [Fe/H]. 
In panel (b), an approximate relation for dynamical $\log g$
given by equation (8) is also depicted by a dashed line.
The results for each star are distinguished by the symbol type: 
SU~Cas --- open circles, DT~Cyg --- filled triangles, SZ~Tau --- open 
triangles, V1334~Cyg --- Greek crosses (+), RT~Aur --- filled squares,
T~Vul --- open downward triangles, FF~Aql --- St. Andrew's crosses 
($\times$), $\delta$~Cep --- filled circles, $\eta$~Aql --- open diamonds,
S~Sge --- filled downward triangles, $\zeta$~Gem --- open squares,
X~Cyg --- filled diamonds.
 }
\label{fig20}
\end{minipage}
\end{figure*}

\setcounter{figure}{20}
\begin{figure*}
\begin{minipage}{120mm}
\includegraphics[width=12.0cm]{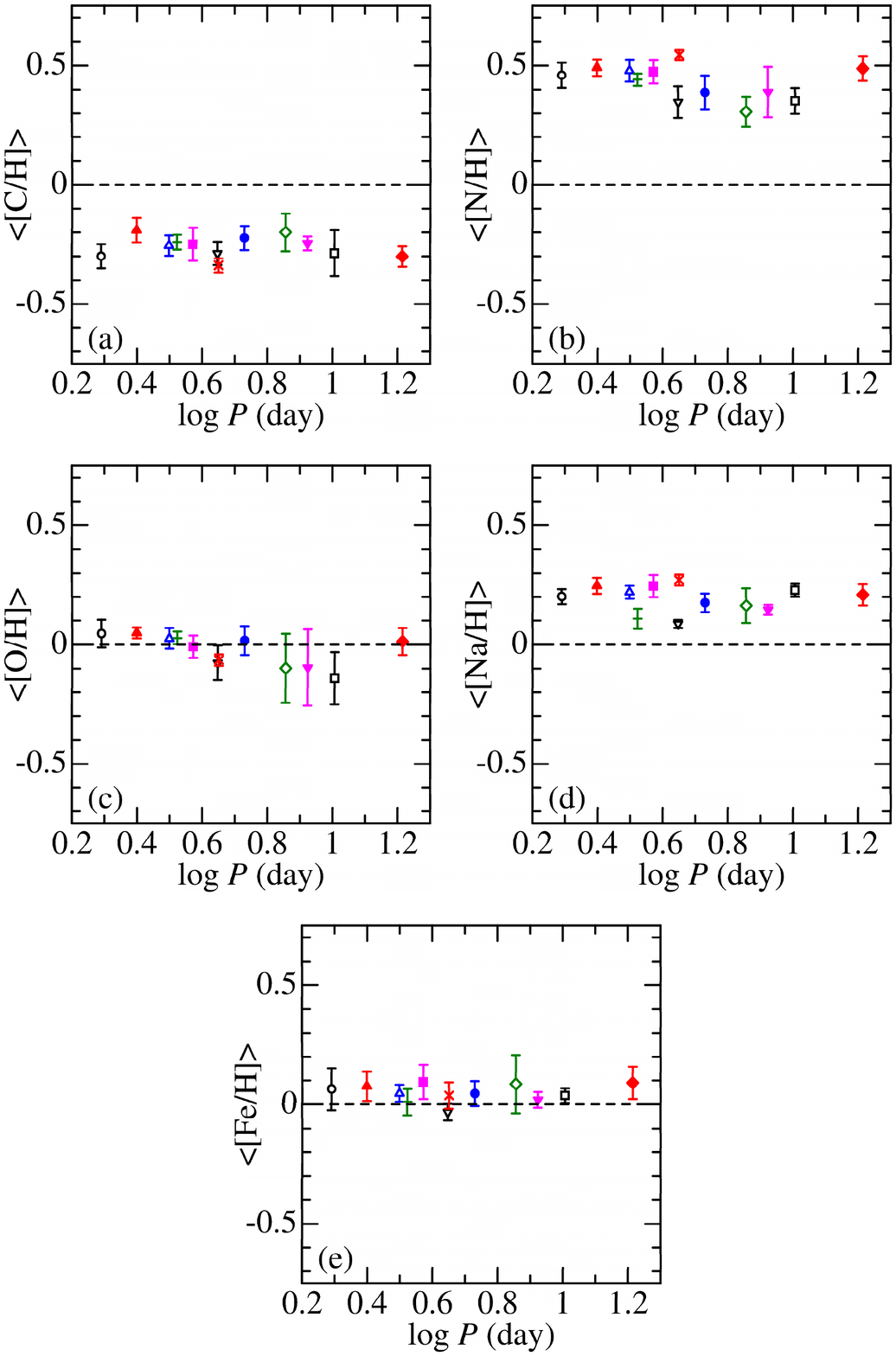}
\caption{$\langle$[X/H]$\rangle$ (mean logarithmic abundance of element X 
relative to the Sun averaged over different pulsation phases) for each 
star, plotted against the pulsation period, based on the data
(expressed in {\it italic}) given in the first line of each section in Table 2:
(a) $\langle$[C/H]$\rangle$, (b) $\langle$[N/H]$\rangle$, 
(c) $\langle$[O/H]$\rangle$, (d) $\langle$[Na/H]$\rangle$, and 
(e) $\langle$[Fe/H]$\rangle$.
The same meanings of the symbols as in Fig. 20, while the attached error 
bar indicate the extent of the standard deviation ($\sigma$).}
\label{fig21}
\end{minipage}
\end{figure*}

\setcounter{figure}{21}
\begin{figure*}
\begin{minipage}{120mm}
\includegraphics[width=12.0cm]{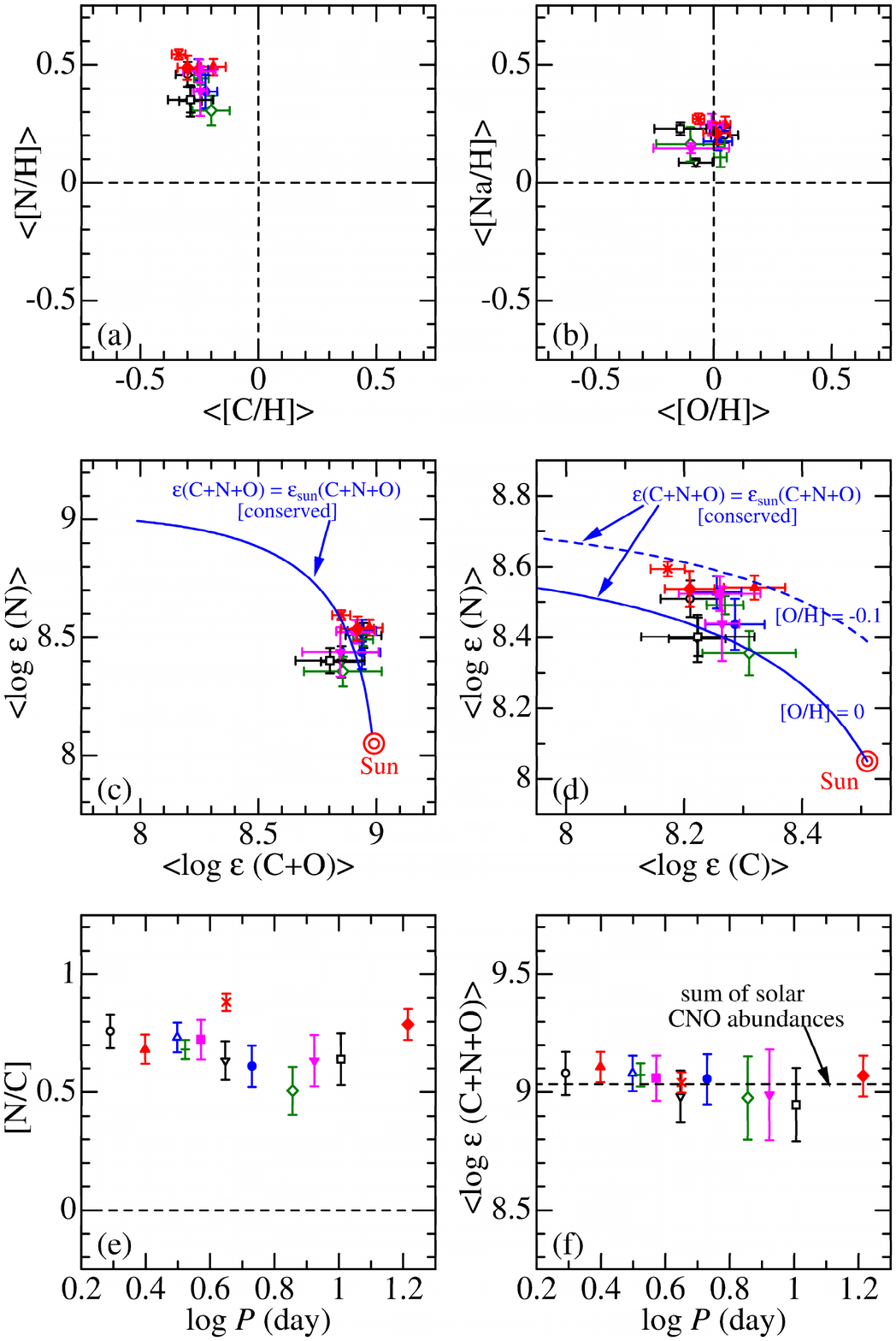}
\caption{Panels (a)--(d) display the correlations between the C, N, O, 
and Na abundances (averaged over different phases): 
(a) $\langle$[N/H]$\rangle$ vs. $\langle$[C/H]$\rangle$, 
(b) $\langle$[Na/H]$\rangle$ vs. $\langle$[O/H]$\rangle$,  
(c) $\langle \log \epsilon $(N)$\rangle$ vs.
$\langle \log \epsilon$(C+O)$\rangle$, 
and (d)  $\langle \log \epsilon$(N)$\rangle$ vs.
$\langle \log \epsilon$(C)$\rangle$ [$\langle \log \epsilon$(X)$\rangle
\equiv \langle$[X/H]$\rangle$ + $\log \epsilon_{\odot}$].
In panels (c) and (d), the expected relation when 
$\epsilon$(C)+$\epsilon$(N)+$\epsilon$(O) is conserved at the solar
value is also shown. The solid and dashed line in panel (d) correspond 
to the cases of [O/H]~=~0.0 and [O/H]~=~$-0.1$, respectively.
Further, the N-to-C abundance ratio ([N/C] $\equiv$ [N/H]$-$[C/H]) and 
the sum of C+N+O abundances ($\langle \log \epsilon$(C+N+O)$\rangle$)  
for each star are plotted against $\log P$ in panels (e) and (f),
respectively. Otherwise, the same as in Fig. 20.}
\label{fig22}
\end{minipage}
\end{figure*}

\end{document}